\documentclass[prd,aps,twocolumn,preprintnumbers,nofootinbib]{revtex4-1}

\usepackage{amsmath}
\usepackage{graphicx}

\usepackage{hyperref} 
\usepackage{mathrsfs}

\usepackage{slashed}
\usepackage{xcolor}

\newcommand{\beq}{\begin{eqnarray}}
\newcommand{\eeq}{\end{eqnarray}}

\newcommand{\nn}{\nonumber}

\DeclareRobustCommand{\Eq}[1]{Eq.~(\ref{#1})}

\DeclareRobustCommand{\eq}[1]{Eq.~(\ref{eq:#1})}
\DeclareRobustCommand{\eqs}[2]{Eqs.~(\ref{eq:#1}) and (\ref{eq:#2})}
\DeclareRobustCommand{\sec}[1]{Sec.~\ref{sec:#1}}

\DeclareRobustCommand{\fig}[1]{Fig.~\ref{fig:#1}}

\DeclareRobustCommand{\app}[1]{App.~\ref{sec:#1}}

\begin{document} 

\preprint{\vbox{\hbox{MIT--CTP 4873}}}

\title{Matching the Quasi Parton Distribution in a Momentum Subtraction Scheme}

\author{Iain W. Stewart}
\author{Yong Zhao \vspace{0.3cm}}
\affiliation{Center for Theoretical Physics, Massachusetts Institute of Technology, Cambridge, MA 02139, USA}

\begin{abstract}

The quasi parton distribution is a spatial correlation of quarks or gluons along the $z$ direction in a moving nucleon which enables direct lattice calculations of parton distribution functions. It can be defined with a nonperturbative renormalization in a regularization independent momentum subtraction scheme (RI/MOM), which can then be perturbatively related to the collinear parton distribution in the $\overline{\text{MS}}$ scheme. Here we carry out a direct matching from the RI/MOM scheme for the quasi-PDF to the $\overline{\text{MS}}$ PDF, 
determining the non-singlet quark matching coefficient at next-to-leading order in perturbation theory.  We find that the RI/MOM matching coefficient is insensitive to the ultraviolet region of convolution integral, exhibits improved perturbative convergence when converting between the quasi-PDF and PDF, and is consistent with a quasi-PDF that vanishes in the unphysical region as the proton momentum $P^z\to \infty$, unlike other schemes.
This direct approach therefore has the potential to improve the accuracy for converting quasi-distribution lattice calculations to collinear distributions.

\end{abstract}

\maketitle

\section{Introduction}

One of the great successes of QCD are factorization theorems, such as those that enable us to make predictions for deep inelastic scattering (DIS) and Drell-Yan processes at hadron colliders~\cite{Collins:1989gx}. These theorems imply that the scattering cross section can be factorized as a convolution of the partonic cross section and a universal parton distribution function (PDF). The former can be calculated analytically in perturbative QCD, thus allowing for a high precision extraction of the latter from cross section data. PDFs are the most basic and important objects for us to obtain information about hadron structure in modern physics. Since PDFs are intrinsic properties of the hadron that include low energy degrees of freedom, they can only be calculated with nonperturbative methods such as lattice QCD. Although their low-order moments have been directly calculated with lattice methods~\cite{Detmold:2001dv,Detmold:2002nf,Dolgov:2002zm}, our most precise knowledge about them comes from global fits to experimental data~\cite{Ball:2014uwa,Dulat:2015mca,Martin:2009iq,Alekhin:2017kpj}  (see~\cite{Buckley:2014ana} for the LHAPDF comparisons).

In parton physics, PDFs are defined as the nucleon matrix elements of light-cone correlation operators. For example, in dimensional regularization with $d=4-2\epsilon$, the bare unpolarized quark distribution function is
\begin{align}
q_i(x,\epsilon) \equiv \!\int\!\! \frac{d\xi^-}{4\pi} \, e^{-ixP^+\xi^-} 
   \! \big\langle P \big| \bar{\psi_i} (\xi^-) \gamma^+ 
   W(\xi^-\!,0)
   \psi_i(0) \big|P\big\rangle ,
  \label{pdf}
\end{align}
where $x$ is the momentum fraction, $i$ is a flavor index, the nucleon momentum $P^\mu=(P^0,0,0,P^z)$, $\xi^\pm = (t\pm z)/\sqrt{2}$ are the light-cone coordinates, and the Wilson line $W$ is given by the path-ordered exponential
\begin{align}
 W(\xi^-,0) &= P \exp\bigg(-ig \int_0^{\xi^-}d\eta^- A^+(\eta^-) \bigg) \,.
\end{align}
The renormalized PDFs $q_j(y,\mu)$ are defined in the $\overline{\text{MS}}$ scheme as
\begin{align}
q_i(x,\epsilon) = \sum_j \int_x^1 \frac{dy}{y} Z^{\overline{\text{MS}}}_{ij}\left( \frac{x}{y}, \epsilon, \mu\right) q_j(y,\mu)\ ,
\end{align}
 $Z^{\overline{\text{MS}}}$ is a function of $x/y$, the flavor indices $i,j$ include the gluon, and $\mu$ is the renormalization scale. The $\overline{\text{MS}}$ definition of the PDF has an interpretation as a parton number density in the light-cone gauge $A^+=0$, and is the most widely used definition for the PDF in factorization theorems. The dependence of the PDF correlator on the light-cone  makes it essentially impossible to directly calculate them using lattice QCD in Euclidean space with imaginary time.

In Ref.~\cite{Ji:2013dva}, Ji proposed that instead of calculating light-cone correlations, one can start from a spatial correlation---called a quasi-PDF---which can be calculated in lattice QCD. The bare quasi-PDF is defined in momentum space ($x$) and coordinate space ($z$) as
\begin{align} \label{eq:quasipdf}
& \tilde{q}(x, P^z, \epsilon) 
  \equiv \int_{-\infty}^\infty \frac{dz}{2\pi}\ e^{ixP^zz} \tilde{q}^B_i(z,P^z,\epsilon) 
  \,,
  \\
& \tilde{q}_i(z,P^z,\epsilon) 
 \equiv \frac12 \big\langle P \big| \bar{\psi}(z) \gamma^z W_z(z,0) \psi(0) \big|P\big\rangle 
  \,, \nonumber
\end{align}
where the spacelike Wilson line is
\begin{align}
 W_z(z,0) &= P \exp\left(-ig\int_0^zdz' A^z(z') \right) \,.
\end{align}
Unlike the PDF in Eq.~(\ref{pdf}) that is invariant under a boost along the $z$ direction, the quasi-PDF changes dynamically under such a boost, which is encoded by its nontrivial dependence on the nucleon momentum $P^z$. The quasi-PDF in coordinate space $\tilde q_j(z,P^z,\tilde \mu)$ is multiplicatively renormalized in dimensional regularization, so we can write 
\begin{align}\label{eq:qPDFposZ}
 \tilde{q}_i(z,P^z\!,\epsilon) = \tilde Z_{i}^X(z,P^z\!,\epsilon,\tilde \mu) \ 
     \tilde q_i^X(z,P^z\!,\tilde \mu) \,.
\end{align}
Here the position space renormalization factors $\tilde Z_{i}^X(z,\epsilon,\tilde \mu)$ are defined in a particular scheme $X$, such as $\overline{\rm MS}$, a momentum-subtraction scheme etc., and $\tilde{\mu}$ is a renormalization scale for the quasi-PDF (whose definition also depends on the scheme $X$).  Fourier transforming to momentum space as in \Eq{eq:quasipdf}, the renormalization for the quasi-PDF involves a convolution in the momentum fraction, 
\begin{align} \label{eq:qPDFmomZ}
\tilde{q}_i(x,P^z\!,\epsilon) 
=\!  P^z\!\! \int_{-\infty}^{+\infty} \!\!\!\!\!\!\!\! dx' \: \tilde{Z}_{i}^X( x\!-\!x', P^z\!,\epsilon, \tilde{\mu}) \, \tilde{q}_i^X(x',P^z\!,\tilde{\mu})  \,.
\end{align}
The structure of the renormalization of the quasi-PDF in \eqs{qPDFposZ}{qPDFmomZ} is similar to that of the quark beam-function~\cite{Stewart:2009yx,Stewart:2010qs}, which is a proton distribution with separations along both the plus and minus light-cone directions. Ref.~\cite{Stewart:2010qs} gives an all orders proof of the position space multiplicative renormalization of the beam function, and this proof also implied that $Z_{ij}\propto \delta_{ij}$, so there is never parton mixing in this case. 
In dimensional regularization it has been explicitly demonstrated that the quasi-PDF is multiplicatively renormalized to two loops~\cite{Ji:2015jwa}. Recently a proof of the multiplicative renormalization has been given both non-perturbatively in Ref.~\cite{Ji:2017oey} and diagrammatically in Ref.~\cite{Ishikawa:2017faj}.  The multiplicative renormalization property present in \eq{qPDFposZ} essentially follows from the known renormalization structure of QCD and of Wilson lines.  Ref.~\cite{Ishikawa:2017faj} has also demonstrated that there is no flavor mixing in the renormalization of the quasi-PDF.

For a nucleon moving with finite but large momentum $P^z\gg \Lambda_{\rm QCD}$, the quasi-PDF can be matched onto the PDF through a momentum space factorization formula~\cite{Ji:2013dva,Ji:2014gla}:\footnote{In this formula we write $\mu/|y|P_z$ for the third argument of the matching coefficient. This has recently been proven to be the correct result in Ref.~\cite{Izubuchi:2018srq}.}
\begin{align} \label{eq:factorization}
\tilde{q}_i^X(x, P^z, \tilde{\mu}) &= \int_{-1}^{+1} \frac{dy}{|y|} \ C_{ij}^X\left(\frac{x}{y}, \frac{\tilde{\mu}}{P^z},\frac{\mu}{|y|P^z}\right) q_j(y,\mu)
\nn\\ 
 &\quad
 + {\cal O}\bigg(\frac{M^2}{P_z^2}, \frac{\Lambda_{\text{QCD}}^2}{P_z^2} \bigg)
  \ ,
\end{align}
where $C_{ij}$ is the matching coefficient, and the ${\cal O}(M^2/ P_z^2, \Lambda_{\text{QCD}}^2/ P_z^2)$ terms are higher-twist corrections suppressed by the nucleon momentum ($M$ is the nucleon mass). Here $q_j(y,\mu)$ for negative $y$ corresponds to the anti-quark contribution. Note that the matching coefficient depends on the quasi-PDF scheme choice $X$, and that for $q_j(y,\mu)$ we always assume the $\overline{\rm MS}$ scheme. Both sides of \eq{factorization} are formally $\mu$ independent, but both do depend on the scale $\tilde \mu$ for the renormalized quasi-PDF, and this dependence need not be small. The indicated power corrections are related to higher-twist contributions in the quasi-PDF. Note that it is important to distinguish between the renormalization of the PDF and quasi-PDF given by the $Z_{ij}$s and $\tilde Z_{ij}$s, and the matching coefficients given by the $C_{ij}$s. The renormalization constants occur in a relation between bare and renormalized matrix elements for the same operators.  On the other hand the matching coefficients occur in a relation between renormalized matrix elements of different operators. The $\tilde{q}$ and $q$ have the same infrared (IR) divergences (which are collinear divergences in Minkowski space), and at perturbative scales $\mu$ and $\tilde\mu$ the $C_{ij}$s can be calculated order by order in $\alpha_s$.  

Based on Ji's proposal, the procedure of calculating PDF from lattice QCD can be summarized as:
\begin{enumerate}
\item Lattice simulation of the quasi-PDF;

\item Renormalization of the quasi-PDF in a particular scheme on the lattice;

\item Subtraction of higher-twist corrections;

\item Matching quasi-PDF in the particular scheme to PDF in the $\overline{\text{MS}}$ scheme.

\end{enumerate}
Efforts have been made to use this proposal to calculate the iso-vector quark distributions $f_{u-d}$, including unpolarized, polarized, and transversity distributions, as well as pion distribution amplitude, from lattice QCD~\cite{Lin:2014zya,Alexandrou:2015rja,Chen:2016utp,Alexandrou:2016jqi,Zhang:2017bzy,Alexandrou:2017huk,Chen:2017mzz,Lin:2017ani}. For this channel the mixing is not important and so the indicies $i$ and $j$ are dropped. The one-loop matching coefficients were first calculated in a transverse momentum cutoff scheme in Ref.~\cite{Xiong:2013bka}, which we denote as $C^{\Lambda_T}(x,\Lambda_T/P^z,\mu/p^z)$. The result was confirmed in Refs.~\cite{Ma:2014jla,Alexandrou:2015rja}. The nucleon-mass corrections of $O(M^{2}/P_z^{2})$ have already been included in the lattice calculations~\cite{Lin:2014zya,Alexandrou:2015rja,Chen:2016utp,Alexandrou:2016jqi,Zhang:2017bzy,Chen:2017mzz,Lin:2017ani}, and the $O(\Lambda^2_\text{QCD}/P_z^{2})$ correction was numerically fitted in Ref.~\cite{Chen:2016utp}. (A direct lattice calculation of the $O(\Lambda^2_\text{QCD}/P_z^{2})$ correction is still desired from the theoretical point of view). In the analyses of Refs~\cite{Lin:2014zya,Alexandrou:2015rja,Chen:2016utp,Alexandrou:2016jqi} the renormalization of the lattice matrix element of quasi-PDF, i.e., Step 2, was absent.  With increasing nucleon momentum $P^z$, lattice renormalization will be a key factor that limits the precision of the calculation of PDFs. Recently, renormalization of the quasi-PDF has been considered Refs~\cite{Ishikawa:2016znu,Chen:2016fxx,Xiong:2017jtn,Constantinou:2017sej,Alexandrou:2017huk,Zhang:2017bzy,Chen:2017mzz,Green:2017xeu,Ji:2017oey,Lin:2017ani}.In other recent work, a related pseudo-PDF distribution was defined~\cite{Radyushkin:2017cyf}, which has been studied in~\cite{Orginos:2017kos}.

One of the standard methods to renormalize operators in lattice QCD is lattice perturbation theory~\cite{Capitani:2002mp}. The perturbative renormalization of the quasi-PDF at one-loop order has recently become available for the Wilson-Clover action~\cite{Xiong:2017jtn,Constantinou:2017sej}.
In practice, it requires a significant amount of work to compute lattice Feynman diagrams for the quasi-PDF,  which limits the ability to go to higher loop orders, and thus the precision that can be achieved. Moreover, fixed-order perturbative renormalization is not reliable when the operator suffers from power divergences under lattice regularization.
An alternative is nonperturbative methods, such as the regularization-invariant momentum subtraction (RI/MOM) scheme, that has been widely used to renormalize local operators on the lattice~\cite{Martinelli:1994ty}. Work in progress to calculate the lattice quasi-PDFs in the RI/MOM scheme has been reported in~\cite{Alexandrou:2017huk,Chen:2017mzz,Green:2017xeu,Lin:2017ani}, and appears to be a promising route for future higher precision quasi-PDF determinations.\footnote{Indeed, the result for the matching coefficient computed in this paper has already been used in the lattice calculation in Ref.~\cite{Chen:2017mzz}.}

In this paper we focus on Step 4 when the lattice quasi-PDF is defined in the RI/MOM scheme. In particular we carry out a perturbative calculation of the matching coefficient that enables this lattice quasi-PDF to be directly matched onto the $\overline{\rm MS}$ PDF.  We denote this matching coefficient by $C^{\text{OM}}$ and it will depend on  the scale of the off-shell subtraction $\mu_R$. The renormalized matrix elements in the RI/MOM scheme are independent of the UV regularization, so we carry out this matching perturbatively with dimensional regularization. Our result for the matching coefficient $C^{\text{OM}}$ also exhibits insensitivity to a UV cutoff in the integral in \eq{factorization}, $|y|< y_{\rm cut}$, unlike the earlier result for $C^{\Lambda_T}$. 

An alternate to the approach we take here would be to convert the lattice quasi-PDF defined with nonperturbative renormalization in the RI/MOM scheme back to the $\overline{\rm MS}$ scheme perturbatively~\cite{Constantinou:2017sej}. Here a matching result $C^{\overline{\rm MS}}$ which converts the $\overline{\rm MS}$ quasi-PDF to the PDF would be used. Our approach is more direct, with only a single step involving a perturbative calculation. Nevertheless it would be interesting to compare both approaches. 

In \sec{rimomscheme} we elaborate on the procedure of implementing the RI/MOM scheme for the quasi-PDF. Then in \sec{results} we calculate the RI/MOM scheme quasi-PDF and the one-loop matching coefficient between this quasi-PDF and the PDF in the $\overline{\text{MS}}$ scheme.  We do this first in Feynman gauge and then in a general covariant gauge. We also quote the known result for the transverse cutoff quasi-PDF for comparison. In \sec{discussion} we analyze these results and give a numerical comparison between the quasi-PDF obtained with the matching coefficients and the PDF.    We conclude in \sec{conclusion}.  A few more detailed discussions are left to appendices, including our implementation of vector current conservation (\app{conservation}), a demonstration that the matching result is independent of the choice of infrared regulator (\app{on-shell}), exploring an alternate RI/MOM scheme (\app{altRIMOM}), and deriving a renormalization group equation for the RI/MOM scheme quasi-PDF (\app{rge}).

\section{Renormalization of Quasi PDF in the RI/MOM Scheme}
\label{sec:rimomscheme}

In the RI/MOM scheme we will define the renormalization constant $\tilde Z^{\rm OM}(z,p_R^z,\Lambda,\mu_R)$ in \eq{qPDFposZ} by imposing a condition on the quasi-PDF evaluated with massless quark states $|p s\rangle$ of momentum $p$ and spin $s$ with $p^2\ne 0$.  We study the iso-vector (non-singlet) case, so that we do not have to consider operator mixing. Here the RI/MOM renormalization condition is
\begin{align} \label{eq:rimom}
& \tilde Z^{\rm OM}(z,p_R^z,\Lambda,\mu_R)^{-1}
 \text{\small $\sum_s$}
 \langle p s| \bar{\psi}(z)\gamma^z W(z,0) \psi(0) | p s\rangle 
 \Big|_{\stackrel{\text{\scriptsize $p^2\!=\!-\!\mu_R^2$}}{\text{\scriptsize $\!\!\!\! p^z\!=p_R^z$}}} 
 \nonumber\\
&= \sum_s \langle ps| \bar{\psi}(z) \gamma^z W(z,0) \psi(0) | ps\rangle \Big|_\text{tree}\nonumber\\
&=4 \, p^z e^{-izp^z}\  \zeta \ \Big|_{p^z=p_R^z}  \,,
\end{align}
where in general $p^\mu=(p^0,0,0,p^z)$, 
\begin{align}
 \zeta=  \frac{1}{4p^z} \sum_s \bar u^s \gamma^z u^s 
\end{align} 
contains the non-singlet spinor factor and the sum over spins $s$, and $\mu_R$ is the renormalization scale. In \eq{rimom} $\Lambda$ denotes the UV cutoff and this definition applies for both lattice QCD calculations where $\Lambda=a^{-1}$ is the inverse lattice spacing, and for continuum dimensional regularization calculations where $\Lambda=\epsilon$. In addition to fixing $p^2=-\mu_R^2$, the choice of the momentum $p^z$ is also a free parameter that is part of what specifies the scheme, and which, for example, does not have to be equal to the proton momentum $P^z$. For this reason we have denoted this dependence in $\tilde Z^{\rm OM}$ by $p_R^z$.  This slightly generalizes \eqs{qPDFmomZ}{factorization} since now there are two arguments $p_R^z$, and the momentum of the state $P^z$ appearing in the quasi-PDF matrix element. The factorization formula for matching in \eq{factorization} with the quasi-PDF in the RI/MOM scheme therefore becomes
\begin{align} \label{eq:factorizationOM}
& \tilde{q}_i^{\,\rm OM}(x, P^z, p_R^z, \mu_R ) 
 \\
  &\quad
  = \int_{-1}^{+1} \frac{dy}{|y|} \ C_{ij}^{\rm OM}\left(\frac{x}{y}, \frac{\mu_R}{p_R^z},\frac{\mu}{|y|P^z}, \frac{yP^z}{p_R^z}\right) q_j(y,\mu)
\,. \nn
\end{align}

On the lattice the renormalization is imposed with a Euclidean momentum, $p^2=-p_E^2$, and we set $p_E^2=\mu_R^2$ in the condition in \eq{rimom}.  For an on-shell vector current matrix element the unique Dirac structure would be $\gamma^z$, but  
since the external state is off-shell, the Feynman diagrams will lead to additional structures such as $p^z\slashed p/p^2$. To deal with this, we evaluate the external states with Dirac spinors $\bar{u}(p,s)$ and $u(p,s)$ and replace
\begin{align} \label{eq:pslash}
\sum_s \bar u(p,s) \Upsilon(z) u(p,s)
& 
  \to \mbox{Tr}\big[ \slashed p\, \Upsilon(z) \big]  \zeta
 \ ,
\end{align}
in order to fully define our off-shell prescription for evaluating the matrix element on the LHS of \eq{rimom}. 

An alternate to using \eq{pslash} would be to reduce the Dirac structure of $\Upsilon$ to a minimal basis of $\{\gamma^z, \slashed{p}\}$ terms, and then only utilize the coefficient of the term proportional to $\gamma^z$ to define the RI/MOM quasi-PDF,
\begin{align}  \label{eq:gammaz}
\sum_s \bar u(p,s) \Upsilon(z) u(p,s)
& \to \Upsilon(z)\Big|_{\gamma^z}\: \zeta \,.
\end{align}
 We will explore this alternate definition of the RI/MOM scheme in Appendix~\ref{sec:altRIMOM}, using the notation $\tilde{q}^{\,{\rm OM}\gamma^z}(x, P^z, p_R^z, \mu_R )$ and $C^{\,{\rm OM}\gamma^z}$.

A key advantage of using the RI/MOM scheme is that although the renormalization factor $\tilde Z^{\rm OM}$ and bare matrix element depend on the choice of regulator, the renormalized quasi-PDF does not,
\begin{align} \label{eq:qrimom}
& \tilde q^{\,\rm OM}(z,P^z,p_R^z,\mu_R) 
 \\
 &= \tilde Z^{\rm OM}(z,p_R^z,\Lambda,\mu_R)^{-1}  
 \langle P| \bar{\psi}(z)\gamma^z W(z,0) \psi(0) | P \rangle
 \nonumber ,
\end{align}
where the state $|P\rangle$ may be a proton for the proton quasi-PDF, or a quark for the quark quasi-PDF, etc. 
Here the $\Lambda$ dependence formally cancels out between $\tilde Z^{\rm OM}$ and the matrix element on the RHS, and the final result for the renormalized quasi-PDF is independent of the choice of UV regulator. 
For this reason  $\tilde q^{\,\rm OM}(z,P^z,p_R^z,\mu_R)$ is referred to as regulator independent.  In practice there may be power suppressed cutoff dependence in the renormalized quasi-PDF due to approximations used for its calculation. For example, on the lattice 
there are discretization effects of ${\cal O}(ap^z, a\sqrt{p_E^2}, a\mu_R)$ which are small in the region $\Lambda_\text{QCD} \ll p^z, \sqrt{p_E^2}, \mu_R \ll 1/a$, and only formally vanish in the continuum limit. These effects can  be reduced to $O(a^2)$ with improved action methods~\cite{Martinelli:1994ty}. 

In the lattice implementation it should be emphasized that the renormalization constant in \eq{rimom} is computed in Euclidean space, but still properly renormalizes the Minkowski space matrix elements. The quasi-PDF proton matrix element in \eq{qrimom} is computed as a Minkowskian matrix element, and hence has the same infrared behavior as the standard PDF, a point that is crucial for the matching in \eq{factorization}. Thus the difference between collinear singularities in Minkowski space and the Euclidean space~\cite{Carlson:2017gpk} does not affect the quasi-PDF paradigm. This issue has been addressed in detail in Refs~\cite{Briceno:2017cpo,Ji:2017rah}.

As has been shown in Ref.~\cite{Xiong:2013bka}, the quasi-PDF has a linear ultraviolet (UV) divergence. The linear divergence arises from the self energy of the finite length space-like Wilson line $W(z,0)$ in the quasi-PDF, which can be renormalized as~\cite{Dotsenko:1979wb,Craigie:1980qs,Dorn:1986dt}
\begin{align}
W^B(z,0) = Z_z e^{\delta m |z|} W^R(z,0)\ ,
\end{align} 
where ``B" and ``R" stand for bare and renormalized. The exponential factor $e^{\delta m |z|}$ introduces counterterms that cancel the linear divergences $\delta m\sim \Lambda$, whereas the rest of the renormalization factor $Z_z$ depends on the end points of the Wilson line, including the coordinates $0$ and $z$, and includes only logarithmic divergences. We can generalize this renormalization relation to gauge-invariant nonlocal quark bilinear operators, as was proven in~\cite{Ji:2017oey,Ishikawa:2017faj}, so that the quasi-PDF renormalization can be split into two parts
\begin{align} \label{eq:multiren}
  \tilde Z(z,p_R^z,\Lambda,\tilde\mu) = e^{\delta m |z|} \tilde Z_\psi(z,p_R^z,\tilde\mu) \,,
\end{align}
where $\delta m$ contains all linear divergences $\propto \Lambda$, while $\tilde Z_\psi(z,p_R^z,\tilde \mu)$ depends on the end points and includes all logarithmic divergences.
Both $\delta m$ and $\tilde Z_\psi$ can in principle depend on the UV cutoff $\Lambda$, renormalization parameter $\tilde\mu$, and momentum of the state used to specify the renormalization scheme $p_R^z$. In order to specify a well defined renormalization scheme with the split in \eq{multiren}, a distinct definition must be given for $\delta m$ and $\tilde Z_\psi$. It would be useful if we can redefine the quasi-PDF to make it free of linear divergence, such as the treatment for transverse momentum distributions~\cite{Collins:2011zzd,Ji:2014hxa}, or the gradient flow method~\cite{Monahan:2016bvm}, but a practical solution on the lattice has not yet been proposed or carried out. With a lattice regulator $\delta m$ will include a linear divergent term $\propto 1/a$, and properly canceling these linear divergences is important numerically. Note that the multiplicative renormalization together with the factorization formula for the renormalized quasi-PDF should enable one to predict the mixed UV-IR two-loop terms discussed in Ref.~\cite{Li:2016amo}.   One strategy is to determine $\delta m$ non-perturbatively from the renormalization of Wilson loop that corresponds to the static quark-antiquark potential~\cite{Musch:2010ka,Ishikawa:2016znu,Zhang:2017bzy,Green:2017xeu}, which has been explicitly verified to cancel the linear divergence in the quasi-PDF at one loop~\cite{Ishikawa:2016znu}.

In the RI/MOM scheme it is not necessary to separately define $\delta m$ and $\tilde Z_\psi$, and we can simply use $\tilde Z^{\rm OM}$ which includes all divergences. Both linear and logarithmic UV cutoff dependence cancel out in \eq{qrimom}, and this cancellation relies only on the multiplicative renormalizability in \eq{qPDFposZ}. This appears to be the simplest approach when carrying out a fully nonperturbative renormalization on the lattice. (If $\tilde Z_\psi(z)$ is calculated in lattice perturbation theory, then it is likely beneficial to separately define and calculate $\delta m$ non-perturbatively.) This also suffices for our purposes for defining the renormalized quasi-PDF, since it is only needed as input for our calculation that matches directly onto the $\overline{\rm MS}$ PDF by determining the coefficient $C^{\rm OM}$ to be used in \eq{factorization}.  We exploit the independence of the RI/MOM scheme to the choice of UV regulator to carry out this matching calculation using dimensional regularization.

\section{Matching between quasi-PDF and PDF}
\label{sec:results}

In this section we calculate the one-loop matching coefficient that converts the renormalized quasi-PDF in the RI/MOM scheme to renormalized PDF in the $\overline{\text{MS}}$ scheme. To do the matching, we compare the two matrix elements in an off-shell quark state with momentum $p^\mu=(p^0,0,0,p^z)$, and $p^2<0$. In dimensional regularization with $d=4-2\epsilon$ and an expansion about $\epsilon=0$, we do not see the linear divergence.  Although the operator is gauge invariant, its off-shell matrix element is generally not. As a result, the renormalization constant defined in the momentum subtraction scheme will be gauge dependent too. We will start by carrying out this calculation in the Feynman gauge. The result in a general covariant gauge will then be given  in Sec.~\ref{landaug}, which includes the Landau-gauge result that is the most relevant for lattice simulations.

\begin{figure*}
\centering
\includegraphics[width=0.8\textwidth]{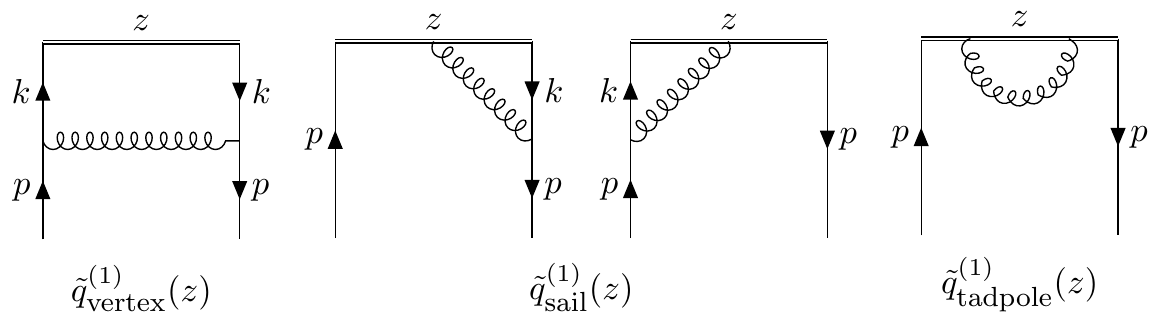}
\caption{One-loop Feynman diagrams for the quasi-PDF. The standard quark self energy wavefunction renormalization is also included, and denoted $\tilde q_{\rm w.fn.}^{(1)}(z)$.}
\label{diagrams}
\end{figure*}
Although in this section we  carry out the matching using the off-shellness as an IR regulator, the result for the matching coefficient is independent of the choice of IR regulator.  Therefore if we carry out the calculation with an on-shell IR regulator like dimensional regularization we will obtain the same matching coefficient between the RI/MOM quasi-PDF and $\overline {\rm MS}$ PDF. To demonstrate this explicitly in \app{on-shell} we repeat the one-loop calculation carried out in this section using dimensional regularization as the IR regulator. 

To define the off-shell quark matrix element we will use the definition in \eq{pslash} (the results obtained for the alternate definition in \eq{gammaz} are given in \app{altRIMOM}). For the quasi-PDF, we use the momentum space Feynman rules for the coordinate space $\tilde q(z,p^z,\epsilon)$~\cite{Ishikawa:2016znu}. At tree level we obtain
\begin{align} \label{eq:tree}
  \tilde q^{(0)}(z,p^z) = 4 p^z \zeta\: e^{-i z p^z} \,.
\end{align}
The one-loop Feynman diagrams are shown in Fig.~\ref{diagrams}.
\begin{widetext}
The displayed diagrams are given by
\begin{align}  \label{eq:qtloopint}
\tilde{q}^{(1)}_\text{vertex}(z,p^z,\epsilon,-p^2) 
&= \zeta\: \mbox{Tr}\left[\slashed p \int {d^d k\over (2\pi)^d}(-ig T^a\gamma^\mu) {i\over \slashed k} \gamma^z {i\over\slashed k} (-i gT^a\gamma^\nu){ - i g_{\mu\nu} \over (p-k)^2}\right]e^{-i k^z z}\ ,
\\
\tilde{q}^{(1)}_\text{sail}(z,p^z,\epsilon,-p^2) 
&= \zeta\: \mbox{Tr}\left[ \slashed p  \int {d^d k\over (2\pi)^d} (ig T^a \gamma^z) {1\over i(p^z-k^z)} \left( 1- e^{-i(k^z-p^z) z}\right)  \delta^{\mu z} {i\over \slashed k} (-ig T^a \gamma^\nu) {-ig_{\mu\nu}\over (p-k)^2}\right] e^{-ip^z z}\nonumber\\
&+ \zeta\: \mbox{Tr}\left[ \slashed p  \int {d^d k\over (2\pi)^d} (-ig T^a \gamma^\nu){i\over \slashed k} (ig T^a \gamma^z) {1\over i(p^z-k^z)} \left( 1- e^{-i(k^z-p^z)z}\right)  \delta^{\mu z}   {-ig_{\mu\nu}\over (p-k)^2}\right] e^{-ip^z z}\ ,
\nn \\
\tilde{q}^{(1)}_\text{tadpole}(z,p^z,\epsilon,-p^2) 
&= \zeta\: \mbox{Tr}\left[ \slashed p  \int {d^d k\over (2\pi)^d} (-g^2)C_F \gamma^z \delta^{\mu z}\delta^{\nu z}\left({1-e^{-i(k^z-p^z) z}\over (p^z-k^z)^2} - {z\over i(p^z-k^z)}\right){-ig_{\mu\nu}\over (p-k)^2} \right]e^{-ip^z z}\ ,
\nn
\end{align}
where in addition we have a contribution from the standard one-loop quark wavefunction renormalization denoted $\tilde q_{\rm w.fn.}^{(1)}(z,p^z,\epsilon,-p^2)$.
\end{widetext}
Here $C_F=4/3$ and $T^a$ is the SU(3) color matrix in the fundamental representation. The second term in the bracket in the last line of \eq{qtloopint}, which is proportional $z$, does not contribute to the loop integral as it is odd under the exchange of $p^z-k^z\to -(p^z-k^z)$.

The quark self-energy correction is $\tilde q_{\rm w.fn.}^{(1)}(z,p^z,\epsilon,-p^2) =\delta Z_\psi\: \tilde q^{(0)}(z,p^z)$ with the tree level matrix element in \eq{tree} and $\delta Z_\psi$ defined in the on-shell scheme to avoid the need for including residue factors when calculating S-matrix elements for the matching. For an off-shell momentum regulator this is
\begin{align} 
\delta Z_\psi = -{\alpha_s C_F\over 4\pi} \left({1\over \epsilon} + \ln {\mu^2 \over -p^2} +1 \right) + O(\alpha_s^2) \, .
\label{onshell}
\end{align}
However, if we follow the definition in \eq{pslash} and integrate over the loop momentum with $z=0$, then contributions from \eq{qtloopint} do not exactly cancel the quark self-energy correction in Eq.~(\ref{onshell}), so the one-loop correction to the conserved local vector current is not zero. This occurs due to the off-shell regulator $p^2\ne 0$, which allows mixing with additional operators due to the off-shellness of the external state.  

When using \eq{pslash}, we can ensure that loop corrections to the vector current cancel, by instead defining $\delta Z_\psi$ in the on-shell scheme using the Ward-Takahashi identity:
\begin{align} \label{ward}
i\Gamma_\mu [p,p] &= {\partial S^{-1}(p) \over \partial p^\mu}\ ,\ \ \delta Z_\psi = {1\over4p^z} \mbox{Tr}\left[\slashed p {\partial \Sigma(p)\over \partial p^z}\right] \,,
 \nn\\
 \delta Z_\psi &= -{\alpha_s C_F\over 4\pi} \left({1\over \epsilon} + \ln {\mu^2 \over -p^2} -1 \right) + O(\alpha_s^2) \, .
\end{align}
Here $\Gamma_\mu[p,p]$ and $S(p)$ are the dressed vertex function and quark propagator, and $\Sigma(p)$ is the quark self energy. For more details see \app{conservation}. Since we work with off-shell quarks for an IR regulator, we will use the latter approach. However, we note that since the $\delta Z_\psi$ contribution is the same for the PDF and quasi-PDF that we will in the end obtain the same result for the matching whether we use Eq.~(\ref{onshell}) or Eq.~(\ref{ward}).

For the matching calculation it is convenient  to use the identity
\begin{align}
f(z) = p^z\!\! \int_{-\infty}^\infty dx\ e^{-ix p^z z} \bigg[ \int_{-\infty}^\infty \frac{dz'}{2\pi}\ e^{ix p^z z'} f(z') \bigg] ,
\end{align}
and carry out the $z'$ integral to express each Feynman diagram as an inverse Fourier transform with respect to the variable $x$. For the quasi-PDF the range of $x$ is unconstrained and we obtain contributions over the full range $-\infty < x < \infty$. 

\begin{widetext}
For example, carrying out the $z'$ integral gives
\begin{align}
\tilde{q}^{(1)}_\text{vertex}(z,p^z,\epsilon,-p^2) 
 + \tilde{q}^{(1)}_\text{w.fn.}(z,p^z,\epsilon,-p^2)
 & = \zeta\:p^z\!\! \int_{-\infty}^\infty\!\!\! dx\: e^{-ix p^z z}\: \mbox{Tr}\left[\slashed p \int {d^d k\over (2\pi)^d}(-ig T^a\gamma^\mu) {i\over \slashed k} \gamma^z {i\over\slashed k} (-ig T^a\gamma^\nu){ - i g_{\mu\nu} \over (p-k)^2}\right] 
  \nn\\
 &\qquad  \times \Big[\delta(k^z-xp^z) - \delta(p^z-xp^z) \Big] \,,
\end{align}
where the difference of $\delta$-functions makes it clear that this contribution vanishes in the local limit $z=0$.  The analogous results for $\tilde{q}^{(1)}_\text{sail}(z,p^z,\epsilon)$ and $\tilde{q}^{(1)}_\text{tadpole}(z,p^z,\epsilon)$ each also contain the same $[\delta(k^z-xp^z) - \delta(p^z-xp^z) ]$ factor. For all contributions we therefore can write
\begin{align}
 p^z\!\!\int\!\! dk^z\! \int_{-\infty}^\infty 
    \!\!\! dx\: e^{-ix p^z z} 
    \Big[\delta(k^z-xp^z) - \delta(p^z-xp^z) \Big] 
 = p^z\!\! \int_{-\infty}^\infty\!\!\! dx\:
    \Big[ e^{-ix p^z z} - e^{-i p^z z} \Big] 
    \int\!\! dk^z\: \delta(k^z-xp^z)
 \,,
\end{align}
and use the $\delta(k^z-x p^z)$ to perform the loop integral over $k^z$.  When evaluating the loop integrals in \eq{qtloopint}, we find that for finite $x$ they are all UV finite and regular as $\epsilon\to 0$. Only the vertex plus wavefunction contribution is UV divergent as $x\to \pm \infty$, but for the RI/MOM scheme these divergences are removed by the counterterm contribution, and hence we can still carry out the calculation with $\epsilon=0$. We denote the bare contributions in this limit by $\tilde{q}(z,p^z,0,p^2)$ where the $p^2$ has been added to indicate the IR regulator. In carrying out the required integrals the following results are useful
\begin{align}
 \int_0^1\!\! dy\: \frac{1}{[(x-y)^2+y(1-y)\rho]^{1/2}} 
   &= \frac{1}{\sqrt{1-\rho}} \ln\bigg( \frac{1-x-\rho/2+\sqrt{1-\rho}|1-x|}{-x+\rho/2+\sqrt{1-\rho}|x|} \bigg)\,,
   \\
 \int_0^1\!\! dy\:  \frac{(1-y)}{[(x-y)^2+y(1-y)\rho]^{3/2}}
  &= \frac{2}{\rho|x|} \frac{(\rho -2|x||1-x|-2x(1-x)}{\rho-4x(1-x)} 
   \,.  \nn
\end{align}
We find that the sum of one-loop contributions,  $\tilde{q}^{(1)} =\tilde{q}^{(1)}_\text{vertex} + \tilde q_{\rm w.fn.}^{(1)} +\tilde{q}^{(1)}_\text{sail} +\tilde{q}^{(1)}_\text{tadpole}$,  for the bare quasi-PDF in Feynman gauge is
\begin{align} \label{finitem}
\tilde{q}^{(1)}(z,p^z,0,-p^2)
&={\alpha_s C_F\over 2\pi}\,(4p^z\zeta)\!\! \int_{-\infty}^\infty \!\!\! dx\,   
 \left(e^{-ixp^z z} - e^{-ip^z z}\right)
 h(x,\rho)
 \,,
\end{align}
where
\begin{align} \label{eq:h}
h(x,\rho) &\equiv
\left\{ \begin{array}{lc}
\displaystyle{1\over \sqrt{1-\rho}}\left[{1+x^2\over 1-x} -{\rho\over2(1-x)}\right] \ln {2x-1+\sqrt{1-\rho}\over 2x-1- \sqrt{1-\rho}} -{\rho\over 4x(x-1)+\rho}+1\ \ \ 
  & x>1\\[10pt]
\displaystyle{1\over \sqrt{1-\rho}}\left[{1+x^2\over 1-x} -{\rho\over2(1-x)}\right] \ln {1+\sqrt{1-\rho} \over 1- \sqrt{1-\rho}} -{2x\over 1-x} \ 
  & 0<x<1\\[10pt]
\displaystyle{1\over \sqrt{1-\rho}} \left[{1+x^2\over 1-x} -{\rho\over2(1-x)}\right] \ln {2x-1-\sqrt{1-\rho}\over 2x-1+ \sqrt{1-\rho}} + {\rho\over 4x(x-1)+\rho}-1\ 
  &x<0
\end{array} \right. 
 \,,
\end{align}
\end{widetext}
and we have defined 
\begin{align}  \label{eq:rho}
 \rho\equiv \frac{(-p^2-i \varepsilon)}{p_z^2} \,,
\end{align}
where the $-i\varepsilon$  allows us to easily analytically continue $\rho$ from $\rho<1$ to $\rho>1$. For $x\to \pm \infty$ the integrand in Eq.~(\ref{finitem}) is $\propto 1/x$ and hence log-divergent (behavior that is cured by the RI/MOM subtraction). On the other hand, the $x$ integral  is convergent at $x=1$. Here the $\exp(-ip^z z)$ term gives a ``real'' contribution with support in $-\infty<x<\infty$, while the $\exp(-ip^z z)$ gives a ``virtual'' contribution proportional to $\delta(1-x)$, and together they provide a well defined result at $x=1$. The one-loop correction to the local vector current is exactly zero as the above integral vanishes at $z=0$. 

By imposing the condition in Eq.~(\ref{eq:rimom}), and writing $Z_{\rm OM}=1+Z_{\rm OM}^{(1)}$ we obtain 
\begin{align}
Z_{\rm OM}^{(1)}(z,p_R^z,0,\mu_R) &= 
  \frac{\tilde{q}^{(1)}(z,p_R^z,0,-p^2=\mu_R^2)}{\tilde{q}^{(0)}(z,p_R^z)}
  \,.
\end{align} 
and the additive counterterm contribution
\begin{align}
\tilde{q}^{(1)}_{\rm{CT}}(z,p^z,p_R^z,\mu_R)
  =- Z_{\rm OM}^{(1)}(z,p_R^z,0,\mu_R)\: \tilde q^{(0)}(z,p^z) \,.
\label{counterterm}
\end{align}
To simplify the presentation of various formulae below we define the dimensionless ratio
\begin{align}  \label{eq:rR}
r_R \equiv \frac{ {\mu_R}^2}{ p_R^{z\,2} }  \,.
\end{align}
In Euclidean space, $p_E^2=p_4^2+p_R^{z\,2} \ge p_z^2$, so the renormalization scale $\mu_R$ one can reach on the lattice by setting $p_E^2=\mu_R^2$ always satisfies $\mu_R^2\ge p_R^{z\,2}$. Therefore, we can consider  $\tilde{q}^{(1)}_{\rm{CT}}$ after analytically continuing to the region $r_R>1$, which is easy to accomplish using the $i\varepsilon$ in \eq{rho}. 

Together these results give the renormalized one-loop quasi-PDF in the RI/MOM scheme
\begin{align} \label{eq:qOM1}
 \tilde q_{\rm OM}^{(1)}(z,p^z,p_R^z,\mu_R)
  &=  \tilde{q}^{(1)}(z,p^z,0,-p^2\ll p_z^{2}) \nn\\
  &\quad + \tilde{q}^{(1)}_{\rm{CT}}(z,p^z,p_R^z,\mu_R) \,.
\end{align}
As indicated, to setup the matching of the quasi-PDF to the PDF, we must keep our physical IR regulator $p^2$ small, i.e. $\rho\ll 1$. Thus we identify the logarithmic IR divergences by Taylor expanding $\tilde{q}^{(1)}$ in $\rho$. At one-loop order there will only be a $\ln\rho$ term, corresponding to the leading logarithmic IR singularity which will also appear in the PDF.  Also, our notation in \eq{qOM1} makes clear that the momentum $p^z$ of the state for which we are considering the quasi-PDF in general need not be equal to the momentum $p_R^z$ that we use for the RI/MOM counterterm. Defining 
\begin{align}
  h_0(x,\rho) &\equiv \left\{
\begin{array}{lc}
\displaystyle {1+x^2\over 1-x}\ln{x\over x-1} +1  \ \ 
 & x>1\\[10pt]
\displaystyle{1+x^2\over 1-x} \ln{4 \over \rho} - \frac{2x}{1-x} \ \
 &0<x<1\\[10pt]
\displaystyle{1+x^2\over 1-x}\ln{x-1\over x} -1 \ \
 &x<0
\end{array}
\right.\,,
\end{align}
the components of the renormalized quasi-PDF in the RI/MOM scheme are 
\begin{widetext}
 \begin{align} \label{eq:renormzpdf1}
&\tilde{q}^{(1)}(z,p^z,0,-p^2\ll p_z^{2})
 = \frac{\alpha_s C_F}{2\pi}\, (4 p^z \zeta)\!\! 
  \int_{-\infty}^\infty\!\! dx\: \left(e^{-ixp^z z} - e^{-ip^z z}\right) 
  h_0(x,\rho) ,
\nn\\
& \tilde{q}^{(1)}_{\rm{CT}}(z,p^z,p_R^z,\mu_R) 
 = -\frac{\alpha_s C_F}{2\pi}\, (4 p^z \zeta)\!\!
  \int_{-\infty}^\infty\!\! dx\: \left(e^{i(1-x)p_R^z z-ip^z z} - e^{-ip^z z}\right) h(x,r_R) 
 \,,
\end{align}
where $h(x,r_R)$ is obtained from \eq{h}.
Here the coupling $\alpha_s=\alpha_s(\mu)$ is taken to be in the standard $\overline{\rm MS}$ scheme. Note that the collinear divergence $\ln\rho$ only appears in the physical region of the PDF $0<x<1$ through $h_0$. The sum can be written as
\begin{align}  \label{eq:renormzpdf}
\tilde q_{\rm OM}^{(1)}(z,p^z,p_R^z,\mu_R)
  &=  \frac{\alpha_s C_F}{2\pi}\, (4 p^z \zeta)
   e^{-izp^z}\! \int\!\! dx \Bigl\{
  \left(e^{i(1-x)p^z z} - 1\right) \bigl[ h_0(x,\rho) - h(x,r_R) \bigr]
  \nn\\
&\qquad\qquad\qquad\qquad\qquad\qquad
   + \left(e^{i(1-x)p^z z} - e^{i(1-x)p_R^z z}  \right) h(x,r_R) 
 \Big\} \,.
\end{align}
For $x \to \pm\infty$ the integrand for the renormalized quasi-PDF behaves as $\propto 1/x^2$ and the integral converges. 
 
The renormalized quasi-PDF in momentum space in the RI/MOM scheme is easily obtained using \eq{renormzpdf}, 	
\begin{align}
\tilde{q}^{(1)}_{\rm{OM}}(x,p^z,p_R^z,\mu_R)  
&=  \int\!\! \frac{dz}{2\pi}\: e^{i x zp^z}\: 
  \tilde q_{\rm OM}^{(1)}(z,p^z,p_R^z,\mu_R)
 \nn\\
&=  \frac{\alpha_s C_F}{2\pi}\, (4 \zeta)
 \biggl\{ \int\!\! dy\: \bigl[ \delta(y-x) - \delta(1-x)\bigr]
  \bigl[ h_0(y,\rho) - h(y,r_R) \bigr] 
  \nn\\
&\qquad\qquad\qquad\qquad
  +  h(x,r_R) - |\eta|\, h\big(1 + \eta(x-1),r_R\big)
   \biggr\}
  \,,
\end{align}
where $\eta \equiv p^z/p_R^z$. 
For the momentum space one-loop quasi-PDF $\tilde{q}^{(1)}_{\rm OM}(x,p^z,\mu_R)$ the difference $\big[\delta(y-x)-\delta(1-x)\big]$ gives a plus-function.  We therefore define the following plus functions with subtractions at $y=1$ 
\begin{align}\label{eq:plus}
\int_1^\infty\!\!\! dy \bigl[{g(y)}\bigr]_\oplus f(y) 
 &=\int_1^\infty\!\!\! dy\ g(y) \big[f(y)-f(1)\big]\,,\nonumber\\
\int_0^1\!\! dy \bigl[ g(y) \bigr]_+ f(y) 
 &= \int_0^1\!\! dy\ g(y) \, \big[f(y)-f(1)\big] \,, 
   \nn\\
\int_{-\infty}^0\!\!\! dy \bigl[ g(y)\bigr]_\ominus \,  f(y) 
&=\int_{-\infty}^0\!\!\!  dy\ g(y) \big[f(y)-f(1)\big] \,,
\end{align}
for arbitrary functions $g(y)$ and $f(y)$.
The renormalized momentum space quasi-PDF in the RI/MOM scheme in Feynman gauge is therefore
\begin{align} \label{renqpdf}
&\tilde{q}^{(1)}_{\rm{OM}}(x,p^z,p_R^z,\mu_R)  \\
&={\alpha_sC_F\over2\pi} (4\zeta)\!
\left\{
\begin{array}{lc}
\bigg[ \displaystyle {1+x^2\over 1-x}\ln{x\over x-1} - {2\over \sqrt{r_R-1}}\left[{1+x^2\over 1-x} -{r_R\over2(1-x)}\right]\arctan {\sqrt{r_R-1}\over 2x-1}+{r_R\over 4x(x-1)+r_R}\bigg]_\oplus \ 
& x>1\\[10pt]
\bigg[ \displaystyle{1+x^2\over 1-x} \ln{4 (p^z)^2\over -p^2} -{2\over \sqrt{r_R-1}}\left[{1+x^2\over 1-x} -{r_R\over2(1-x)}\right] \arctan\sqrt{r_R-1} \bigg]_+ \ 
&0<x<1\\[10pt]
\bigg[ \displaystyle{1+x^2\over 1-x}\ln{x-1\over x}+ {2\over \sqrt{r_R-1}}\left[{1+x^2\over 1-x} -{r_R\over2(1-x)}\right]\arctan {\sqrt{r_R-1}\over 2x-1} - {r_R\over 4x(x-1)+r_R} \bigg]_\ominus \ 
&x<0
\end{array}
\right. \nonumber \\
& \ \ + {\alpha_sC_F\over2\pi} (4\zeta) \bigg\{ 
  h(x,r_R) - |\eta|\, h\big(1 + \eta(x-1),r_R\big)
   \biggr\}
 \nn\,.
\end{align}
Note that the last term vanishes for $\eta=1$.
 
Next we consider the PDF calculated with the same off-shellness regulator, once again using the same identity in \eq{pslash} to uniquely define our treatment of the spinors when working off-shell. With this definition the renormalized one-loop matrix element of PDF in the $\overline{\text{MS}}$ scheme is given by
\begin{align}
q^{(1)}(x,\mu)&= {\alpha_sC_F\over 2\pi} (4\zeta) \left\{
\begin{array}{cc}
\displaystyle 0\ & x>1\\
\bigg[ \displaystyle {1+x^2\over 1-x} \ln\frac{\mu^2}{-p^2} -{1+x^2\over 1-x}\ln\big[x(1-x)\big]-(2-x) \bigg]_+ 
  \qquad  &0<x<1\\
\displaystyle 0\ & x<0
\end{array}\right. 
  \,.
\end{align}
Considering the factorization formula in \eq{factorizationOM},  the matching coefficient $C^{\rm OM}$ between the renormalized quasi-PDF in the RI/MOM scheme and standard PDF in the $\overline{\rm MS}$ scheme is then determined by the difference between the momentum space quasi-PDF and PDF results
\begin{align}
  & C^{\text{OM}}\left(\xi, {\mu_R\over p_R^z},{\mu\over p^z},
  {p^z\over p_R^z}\right)
  = \delta(1-\xi)
  + \frac{1}{4\zeta} \Big[ 
  \tilde{q}^{(1)}_{\rm{OM}}\left(\xi,p^z,p_R^z,\mu_R\right) - q^{(1)}\left(\xi,\mu\right)
  \Big] + {\cal O}(\alpha_s^2) 
   \,,
\end{align}
which gives the one-loop matching coefficient
\begin{align} \label{eq:crimom}
& C^{\text{OM}}\left(\xi, {\mu_R\over p_R^z},{\mu\over p^z},
  {p^z\over p_R^z}\right)
  - \delta(1-\xi) 
  \\[8pt]
&={\alpha_sC_F\over 2\pi}\left\{\! \begin{array}{lc}
\displaystyle \bigg[ {1+\xi^2 \over 1-\xi}\ln{\xi\over \xi-1} 
  -{2(1+\xi^2)-r_R\over (1-\xi)\sqrt{r_R-1}} \arctan {\sqrt{r_R-1}\over 2\xi-1}
  +{r_R\over 4\xi(\xi-1)+r_R} 
 \bigg]_\oplus  
  &  \xi>1\\[15pt]
\bigg[ 
 \displaystyle{1+\xi^2\over 1-\xi} \ln{4 (p^z)^2 \over \mu^2 }+{1+\xi^2\over 1-\xi}\ln{\big[\xi(1-\xi)\big]}+(2-\xi) 
 -{2\arctan\sqrt{r_R-1}\over \sqrt{r_R-1}} \bigg\{ {1+\xi^2\over 1-\xi} -{r_R\over 2(1-\xi)} \bigg\}
 \bigg]_+    &0<\xi<1 \\[15pt]
\bigg[ \displaystyle{1+\xi^2\over 1-\xi}\ln{\xi-1\over \xi}+ {2\over \sqrt{r_R-1}}\left[{1+\xi^2\over 1-\xi} -{r_R\over 2(1-\xi)}\right]\arctan {\sqrt{r_R-1}\over 2\xi-1} - {r_R\over 4\xi(\xi-1)+r_R} \bigg]_\ominus  
  &\xi<0
\end{array} \right. \nonumber \\
& \ \ + {\alpha_sC_F\over2\pi}  \bigg\{ 
  h(\xi,r_R) - |\eta|\, h\big(1 + \eta (\xi-1),r_R\big)
   \biggr\} \,,\nn
\end{align}
where $r_R$ is given in \eq{rR} and $h$ in \eq{h}. When this result is used in the matching formula \eq{factorizationOM} one must take $p^z=yP^z$ and hence $\eta=yP^z/p_R^z$.  This is the result for the matching between quasi-PDF in the RI/MOM scheme and PDF in the $\overline{\text{MS}}$ scheme for the non-singlet case in Feynman gauge.   As expected, $C_\text{OM}$ is independent of the IR regulator $-p^2$ since the logarithmic IR singularities cancel between the quasi-PDF and PDF.  An alternate derivation of \eq{crimom} with a different choice for the IR regulator is given in \app{on-shell}

The result in the RI/MOM scheme in \eq{crimom} also exhibits convergent behavior as $\xi\to \pm\infty$. In section~\ref{sec:compareschemes} below we compare this behavior with the results obtained with the quasi-PDF in a $\overline{\rm MS}$ and transverse cutoff schemes.

\end{widetext}

\subsection{Landau and General Covariant Gauge}
\label{landaug}
In lattice QCD, the RI/MOM scheme is most easily implemented in the Landau gauge, so in this section we extend our calculation to give the matching result for a general covariant gauge where the gluon propagator is
\begin{equation} \label{eq:xigauge}
iD_\tau^{\mu\nu}(k) = -{i\over k^2} \left[ g^{\mu\nu} - (1-\tau) {k^\mu k^\nu\over k^2}\right] \,.
\end{equation}
Here $\tau=0$ corresponds to the Landau gauge. We write the quasi-PDF in a general covariant gauge as the $\tau=1$ Feynman gauge result plus a correction
\begin{align}  \label{eq:covrt}
  \tilde q_{\tau}^{\,\rm OM}(x,p^z,p_R^z,\mu_R) 
   &= \tilde q^{\,\rm OM}(x,p^z,p_R^z,\mu_R)  
   \\
 &\ \ + \Delta\tilde{q}_{\tau}^{\,\rm OM}(x,p^z,p_R^z,\mu_R) 
   \,. \nn
\end{align}
The correction from the $(1-\tau)$ term in \eq{xigauge} is at one-loop given by
\begin{widetext}

\begin{align}
\Delta\tilde{q}_{\tau}^{\,\rm OM}(x,p^z,p_R^z,\mu_R)
 &=(1-\tau) {\alpha_s C_F\over 2\pi}
 (4\zeta) \left\{ \begin{array}{lc}
\displaystyle  \Bigl[ \Delta h_{\tau}(x,r_R) \Bigr]_\oplus 
 \quad & x>1\\[15pt]
0  \ & 0<x<1\\[6pt]
\displaystyle \Bigl[ \Delta h_{\tau}(x,r_R) \Bigr]_\ominus
 & x<0 \end{array} \right.
 \nn\\
 &\ \ + (1-\tau) {\alpha_s C_F\over 2\pi}
 (4\zeta) \Big[
  -\Delta h_{\tau}(x,r_R) + |\eta|\, \Delta h_{\tau}\bigl(1+\eta(x-1),r_R\bigr) \Big]
 \,,
\end{align}
where
\begin{align}
  \Delta h_{\tau}(x,r_R) & \equiv 
\left\{ \begin{array}{lc}
\displaystyle   \frac{(1-2x)\, r_R^2}{ 2(1-x) \left[r_R+4x(x-1)\right]^2} 
 \quad & x>1\\[15pt]
\displaystyle \frac{1-2x}{2(1-x)}\ & 0<x<1\\[6pt]
\displaystyle  \frac{-(1-2x)\, r_R^2}{ 2(1-x) \left[r_R+4x(x-1)\right]^2} 
 & x<0 \end{array} \right.
  \,.
\end{align}
For the PDF with an off-shell momentum regulator, there is also an additional contribution,
\begin{align}
q^{(1)}_\tau(x,\mu)=(1-\tau){\alpha_sC_F\over 2\pi}(4\zeta) 
\left\{ \begin{array}{lc}
\displaystyle 0\ & x>1\\
\displaystyle \bigg( {1-2x\over 2(1-x)} \bigg)_+ \quad & 0<x<1\\
\displaystyle 0\ & x<0 \end{array} \right.
  \ .
\end{align}
As a result, the matching coefficient for a general covariant gauge is given by the Feynman gauge result from \eq{crimom} plus an additional term
\begin{align}
C_\tau^{\text{OM}}\Big(\xi, {\mu_R\over p_R^z},{\mu\over p^z},{p^z\over p_R^z}\Big)
 &= C^{\text{OM}}\Big(\xi, {\mu_R\over p_R^z},{\mu\over p^z},{p^z\over p_R^z} \Big) + (1-\tau){\alpha_sC_F\over 2\pi}
   \left\{ \begin{array}{lc} 
  \Bigl[ \Delta h_{\tau}(\xi,r_R) \Bigr]_\oplus
   & \xi>1\\
\displaystyle \biggl( -{(1-2\xi)\over 2(1-\xi)}\biggr)_+  \ \ \ 
   & 0<\xi<1\\
  \Bigl[ \Delta h_{\tau}(\xi,r_R) \Bigr]_\ominus
   & \xi<0 
\end{array} \right.
\nn\\
 &\ \ + (1-\tau) {\alpha_s C_F\over 2\pi}
 (4\zeta) \Big[
  -\Delta h_{\tau}(\xi,r_R) + |\eta|\, \Delta h_{\tau}\bigl(1+\eta(\xi-1),r_R\bigr) \Big]
\,.
\label{eq:landau}
\end{align}
When utilizing the matching equation with the RI/MOM scheme in a chosen gauge, we note that the RI/MOM quasi-PDF $\tilde{q}_\tau^{\,\rm OM}$ is gauge dependent, as is the matching coefficient $C_\tau^{\rm OM}$, and in both cases this is induced by the presence of the gauge dependent RI/MOM UV counterterm. Therefore this gauge dependence is the same and yields a gauge invariant result for the $\overline{\rm MS}$ PDF order by order in $\alpha_s$. When the quasi-PDF is renormalized non-perturbatively and the matching is carried out perturbatively, then the cancellation will be incomplete, and it would be reasonable for example to look at the residual gauge dependence as a means of assessing an uncertainty from higher orders in perturbation theory. However we will see in \sec{discussion} that at one-loop the gauge dependent terms are much smaller than the residual scale dependence, and hence this is unlikely to be a significant source of uncertainty. For our numerical analysis in \sec{discussion} we will consider both the Feynman gauge result from \eq{crimom}, denoted $C^{\rm OM}$, and the Landau gauge result obtained from \eq{landau} with $\tau=0$, and denoted $C^{\rm OM}_{\tau=0}$. 

\end{widetext}

\subsection{Comparison to Other Schemes}
\label{sec:compareschemes}

The one-loop matching coefficient between the quasi-PDF and PDF was originally calculated in Ref.~\cite{Xiong:2013bka} in an on-shell scheme with the UV divergence regulated by a finite transverse momentum cutoff $\Lambda_T$, using Feynman gauge. Using our notation for the plus functions the result for this scheme is
\begin{widetext}
\begin{align}  \label{eq:CLambdaT}
 C^{\Lambda_T}\Big(\xi,{\mu\over p^z}, \frac{\Lambda}{P^z}\Big)
&= \delta(1-\xi) 
  \\
 &+{\alpha_sC_F\over 2\pi}\left\{\begin{array}{cl}
\bigg[ 
\displaystyle {1+\xi^2\over 1-\xi}\ln{\xi\over \xi-1}+1 + {1\over (1-\xi)^2}{\Lambda_T\over P^z} \bigg]_\oplus  
 & \xi>1\\[10pt]
\bigg[ \displaystyle{1+\xi^2\over 1-\xi} \ln{4 (p^z)^2 \over \mu^2 }
 +{1+\xi^2\over 1-\xi}\ln{\xi(1-\xi)}+1-{2\xi\over 1-\xi} +{1\over (1-\xi)^2} {\Lambda_T\over P^z}\bigg]_+ 
 \quad &0<\xi<1  \\[10pt]
\bigg[ \displaystyle{1+\xi^2\over 1-\xi}\ln{\xi-1\over \xi}-1+{1\over (1-\xi)^2} {\Lambda_T\over P^z}\bigg]_\ominus &\xi<0
\end{array}\right. 
 \,. \nn
\end{align}
\end{widetext}
This  result was used in the lattice calculations of $f_{u-d}$ in Refs.~\cite{Lin:2014zya,Alexandrou:2015rja,Chen:2016utp,Alexandrou:2016jqi}.
Note that the linear divergence is not subtracted in the quasi-PDF in this scheme, so there is no renormalization scale $\mu_R$ associated with it. 

In the recent works~\cite{Constantinou:2017sej,Alexandrou:2017huk}, the quasi-PDF is renormalized in the RI/MOM scheme and matched to the quasi-PDF in the $\overline{\text{MS}}$ scheme. Eventually, the quasi-PDF in the $\overline{\text{MS}}$ scheme needs to be matched to PDF in the $\overline{\text{MS}}$ scheme.  The result for $C^{\overline{\text{MS}}}$ is gauge invariant, which follows because on-shell definitions of the quasi-PDF and PDF in the $\overline{\rm MS}$ scheme are gauge invariant (or alternatively because any gauge dependence associated with off-shell regulation of the infrared physics will be identical for the quasi-PDF and PDF). To carry out the appropriate matching coefficient for this case one must carefully treat UV divergences that come from $x\to \pm \infty$, which implies that there is not a single overall plus function for each region. The result for this case is presented in Ref.~\cite{Izubuchi:2018srq}, and our result in RIMOM is consistent with the scheme conversion formulas presented there.

\begin{figure*}[t!]
	\centering
	\includegraphics[width=0.49\textwidth]{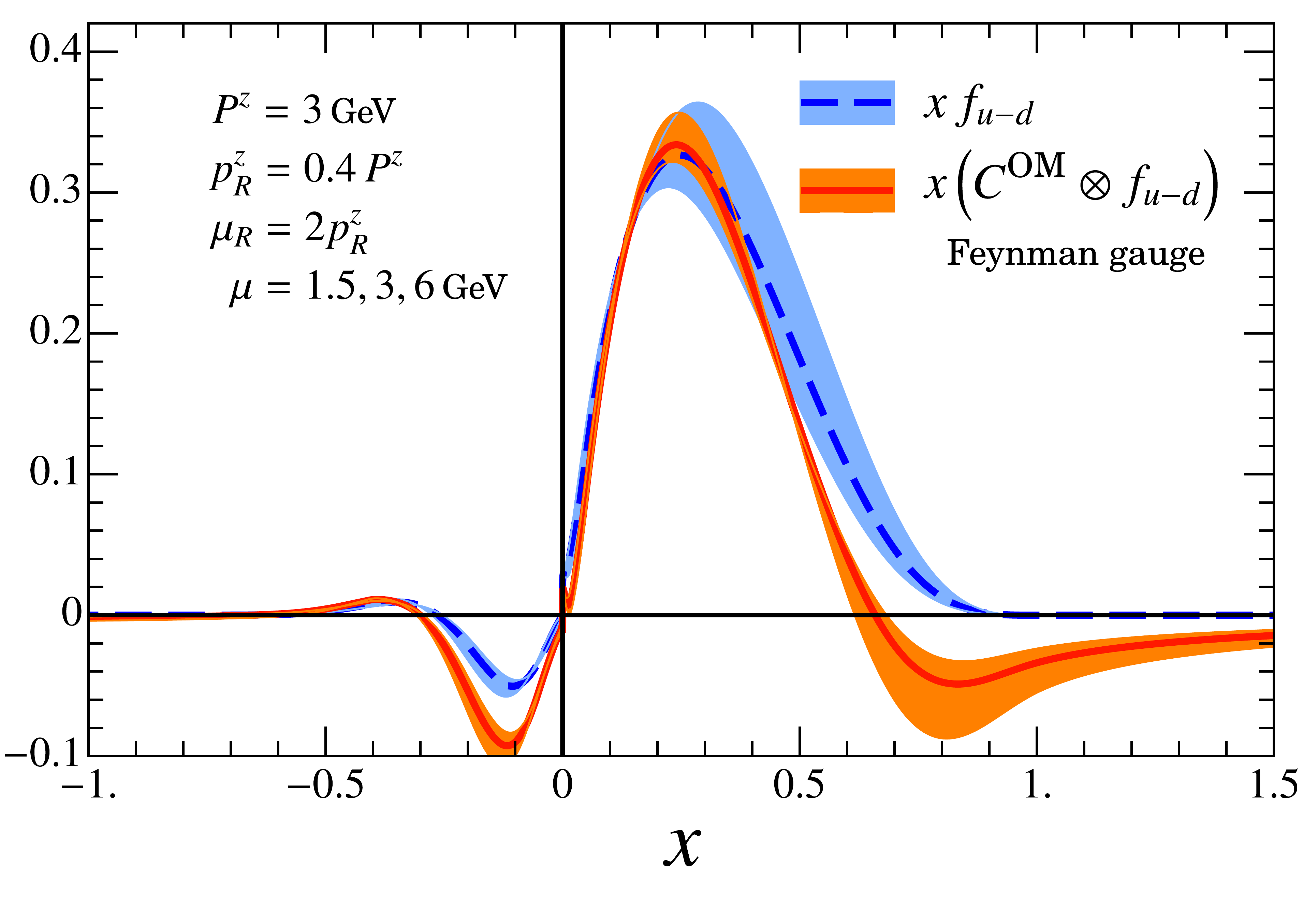}
	\hspace{0.1cm}
	\includegraphics[width=0.49\textwidth]{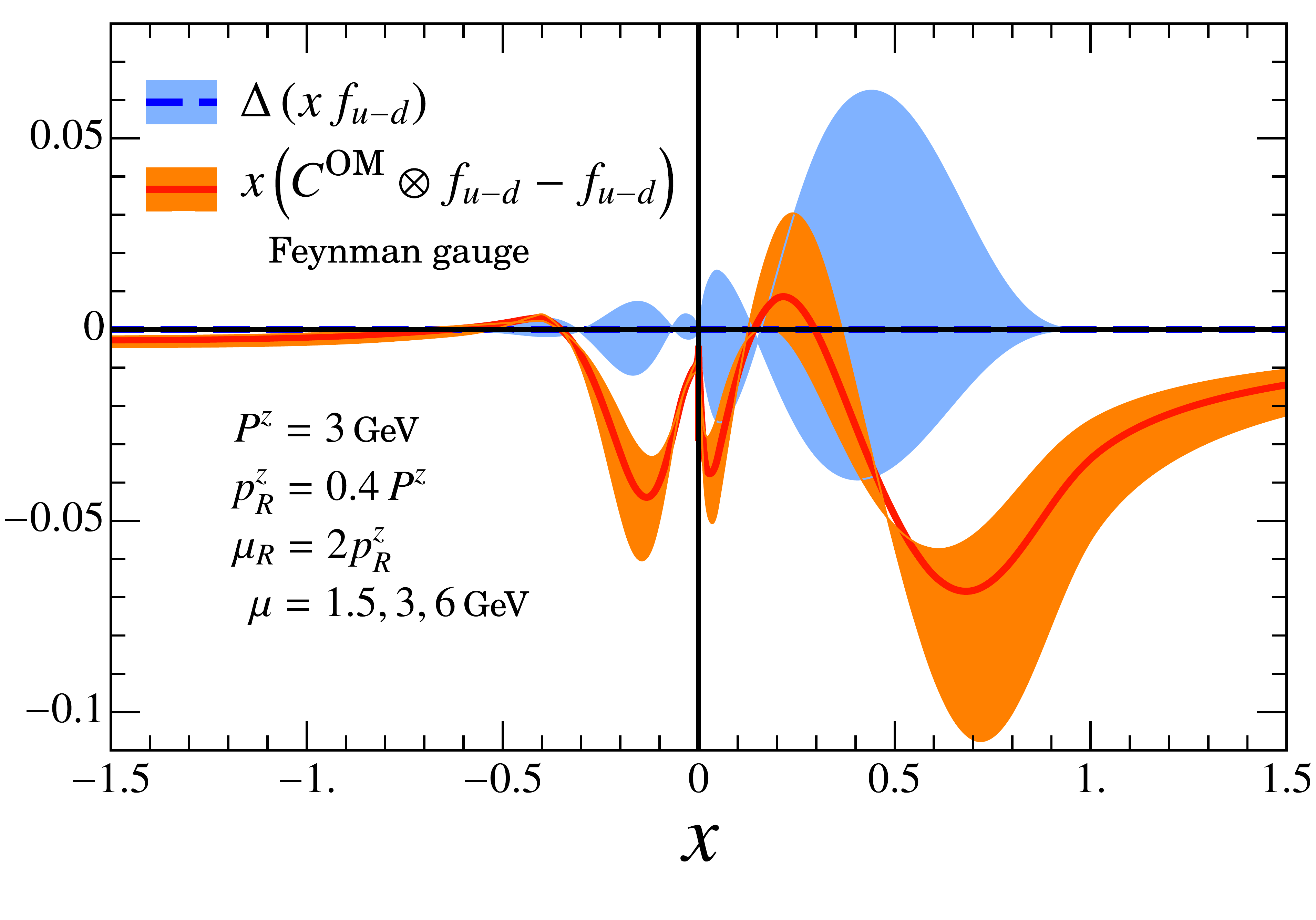}
	\caption{Comparison between the PDF $xf_{u-d}$ and the quasi-PDF result obtained from $x(C^{\text{OM}}\otimes f_{u-d})$ in Feynman gauge. The orange and blue bands indicate the results from varying the factorization scale $\mu$ by a factor of two. Left: $x(C^{\text{OM}}\otimes f_{u-d})$ and $xf_{u-d}$. Right: differences when taking $x(C^{\text{OM}}\otimes f_{u-d})$ or $xf_{u-d}$, and subtracting $xf_{u-d}(x,3\text{ GeV})$.}
	\label{fig:muf}
\end{figure*}

\section{Numerical Analysis}
\label{sec:discussion}

In this section we numerically analyze the quasi-PDF by studying how the matching coefficient in Eq.~(\ref{eq:factorization}) changes the PDF. To calculate the convolution between the momentum space matching coefficient $C^{\text{OM}}(x/y)$ and a PDF $f(y)$ we note that for $-1<y<1$ and fixed $x$ the variable $x/y$ goes over the range $-\infty < x/y < -|x|$ and $|x| <x/y<\infty$. For any plus function $g^{\rm plus}(x/y)$ of the types defined in \eq{plus}, namely with a subtraction at $1$, we can carry out the integral by imposing a soft cutoff $|x/y-1|>\beta=10^{-m}$ on its pure function part, and calculating the coefficient of $\delta$-function which will also depend on $\beta$. When $g^{\rm plus}(x/y)$ is convoluted with an arbitray function, the result should be independent of the soft cutoff $\beta$, as we will check in the calculations.  Alternatively, we can also calculate the convolution by using
\begin{align}  \label{eq:applyplus}
&\int dy\, g^{\rm plus} \Big({x\over y}\Big) \frac{f(y)}{|y|} 
 \\
&= \int dy'\, g_r^{\rm plus}(y') \, \bigg[ \frac{x}{y^{\prime 2}} \,   
  \frac{f(x/y')}{|x/y'|} - x\, \frac{f(x)}{|x|} \bigg]
  \nn\\
&= \int_{-\infty}^\infty dy \left[
 {1\over |y|} g_r^{\rm plus}\Big({x\over y}\Big)f(y) - 
 {1\over |x|} g_r^{\rm plus}\Big({y\over x}\Big)f(x) 
 \right] \,. \nn
\end{align}
Note that here we absorbed the limits $\theta(1-|y|)$ in the function $f(y)$.
Here the subscript $r$ on $g_r^{\rm plus}$ denotes the pure function, which are the argument of the plus functions. For the first equality in \eq{applyplus} we changed variable to $y'=x/y$ and then applied the plus functions. For the last equality in \eq{applyplus} we changed variable back to $y=x/y'$ in the first term, and to $y'=y/x$ in the second term. We have checked that these two methods give the same result.

As an example we use for our analysis the unpolarized iso-vector parton distribution,
\begin{align}
f_{u-d}(x,\mu) =  f_u(x,\mu) - f_d(x,\mu) - f_{\bar{u}}(-x,\mu) +  f_{\bar{d}}(-x,\mu),
\end{align}
where we include 
$f_{\bar{u}}(-x,\mu) = - f_{\bar{u}}(x,\mu)$ and $f_{\bar{d}}(-x,\mu) = - f_{\bar{d}}(x,\mu)$, the anti-parton distributions. We use the next-to-leading-order iso-vector PDF $f_{u-d}$ from ``MSTW 2008"~\cite{Martin:2009iq} with the corresponding running coupling $\alpha_s(\mu)$. For the numerical calculation we impose a UV cutoff on the $y$-integral so that  $|y|< y_{\text{cut}}=10^n$ for any $x$ and some $n>1$. Results in the RI/MOM scheme are independent of this cutoff, whereas we will show below that the transverse cutoff scheme exhibits sensitivity to $y_{\rm cut}$.  We also test soft cutoffs $|y|>10^{-k}$ and $|x/y-1|>10^{-m}$ with $m\ge3$, but find that the results in all schemes are independent of $k$ and $m$. 

As default values for our figures we take $P^z=3.0$ GeV, use the Feynman gauge RI/MOM matching result $C^{\rm OM}$ from \eq{crimom} and the $\overline{\rm MS}$ renormalization scale $\mu=3.0\,{\rm GeV}$. The RI/MOM matching coefficient $C^{\rm OM}$ is a function of $\mu$ and $P^z$, as well as the ratios $P^z/p_R^z$ and $r_R=\mu_R^2/(p_R^z)^2$. The size of the correction induced by the matching coefficient does depend on the values of $P^z/p_R^z$ and $r_R$, and we will elaborate on this below.
For our study, we set the default values $p_R^z=0.4P^z$ and $\mu_R=2.0p_R^z$ and then consider variations of $\mu$, $P^z$, $P^z/p_R^z$, and $r_R$ about these choices.

\begin{figure*}
	\centering
	\includegraphics[width=.49\textwidth]{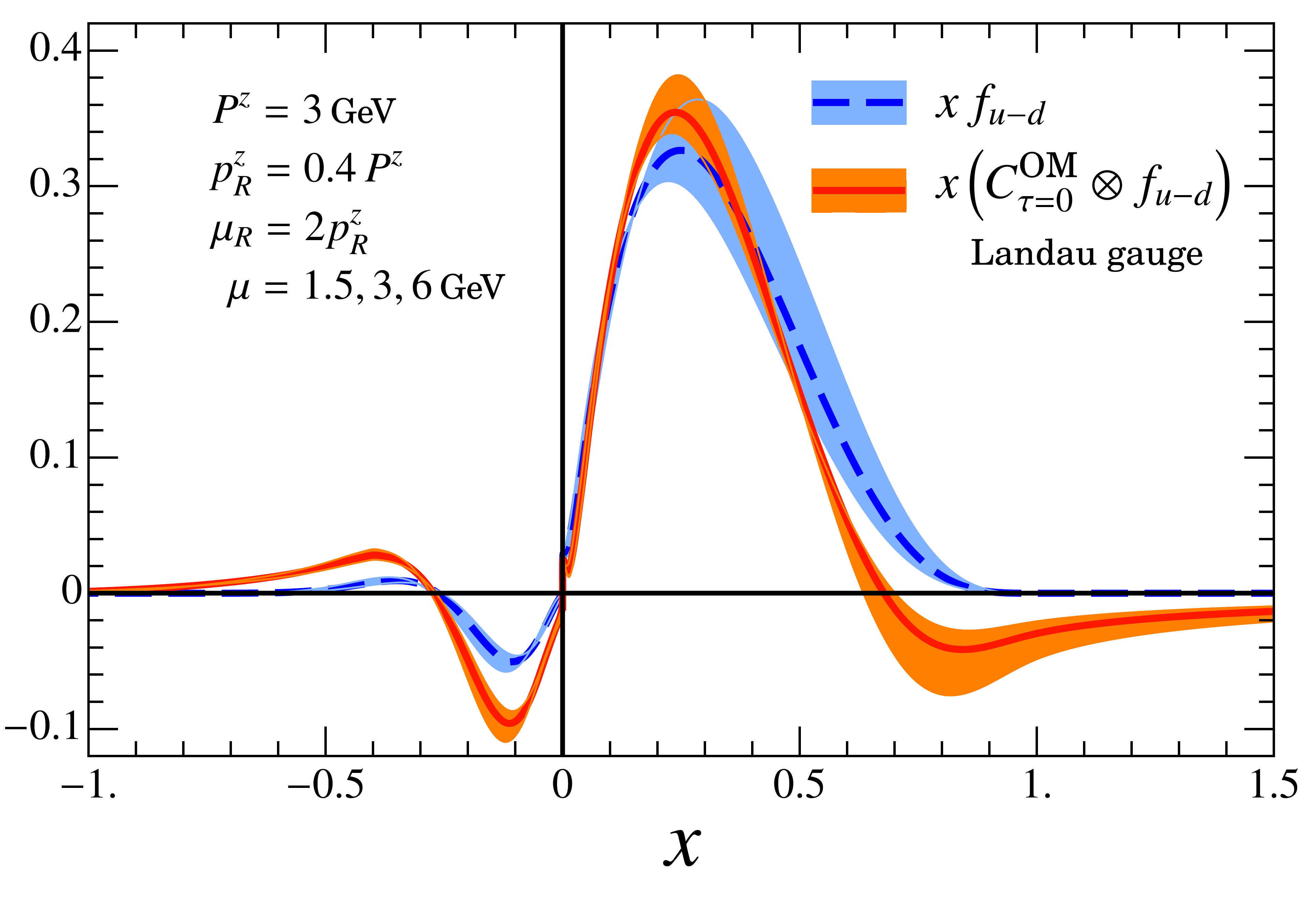}
	\hspace{0.1cm}
    \includegraphics[width=0.45\textwidth]{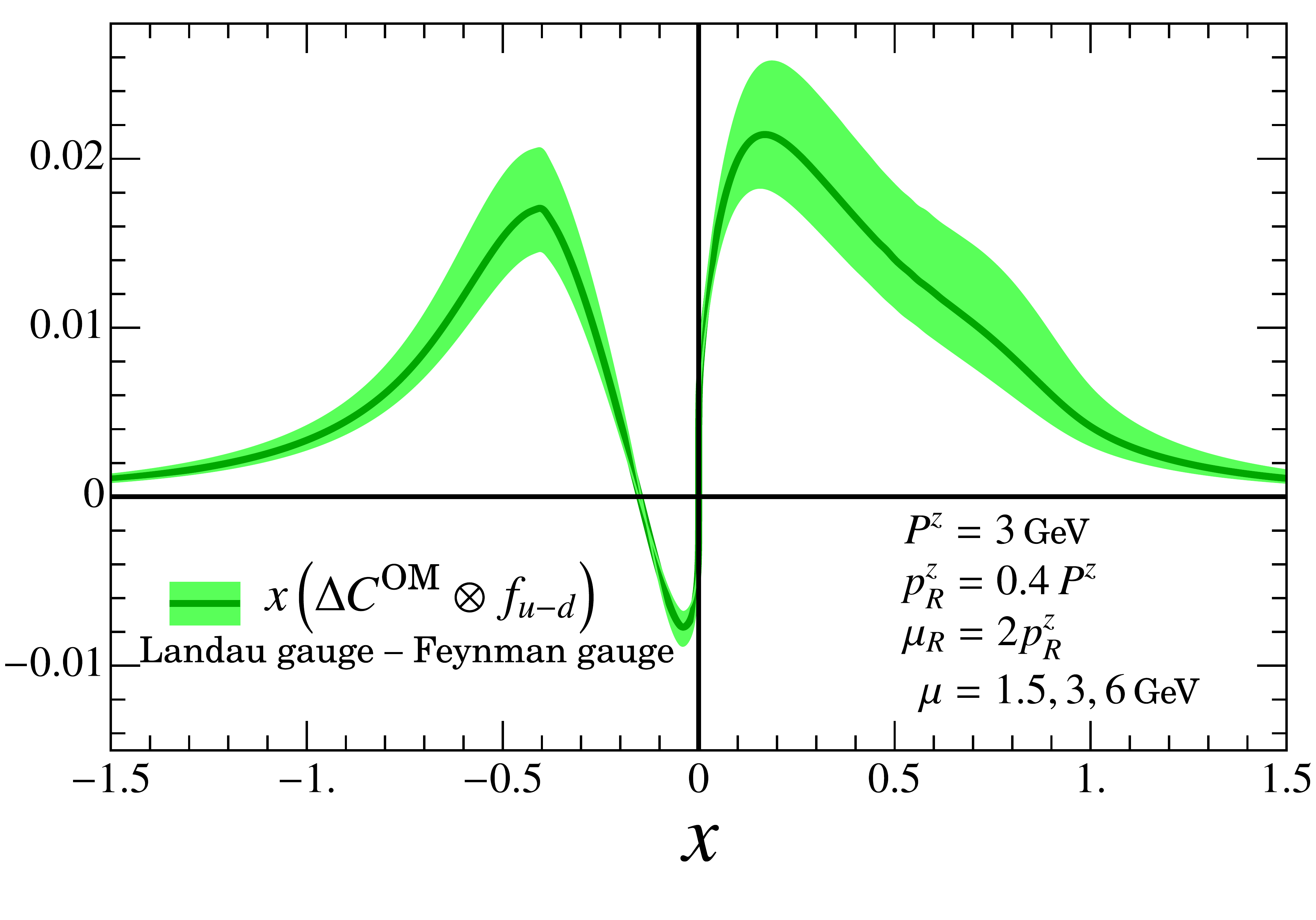}	
	\caption{Comparison between the PDF $xf_{u-d}$ and the quasi-PDF obtained from $x(C^{\text{OM}}\otimes f_{u-d})$ in the Landau gauge. The orange, blue, and green bands indicate the results from varying the factorization scale $\mu$ by a factor of two.}
	\label{fig:landaugauge}
\end{figure*}

First consider Fig.~\ref{fig:muf} which shows a comparison of the RI/MOM quasi-PDF $\tilde q^{\,\rm OM}(x,P^z,p_R^z,\mu_R)=(C^{\text{OM}}\otimes f_{u-d})$ (red solid line) and the $\overline{\rm MS}$ PDF $f_{u-d}(x,\mu)$ (blue dashed line).  In this figure and in others below we multiply by $x$ in order to more easily observe the small $x$ region. The left panel shows the direct comparison, and the right panel makes the comparison subtracting $x f_{u-d}(x,\mu)$. We see that the quasi-PDF and PDF are close to one another, which is appealing for the convergence of perturbation theory. In Fig.~\ref{fig:muf} we also vary the factorization scale $\mu$ by a factor of two, from $\mu=1.5\,{\rm GeV}$ to $\mu=6\,{\rm GeV}$, showing the result by the blue and orange bands about the PDF and quasi-PDF respectively.    For the quasi-PDF from \eq{crimom} the  dependence on $\mu$ cancels out between $C$ and $q$, order by order in perturbation theory, whereas the PDF has a dependence on $\mu$ at leading-logarithmic order, so in the figure a decrease in the $\mu$ dependence is observed as expected. 
As shown in Fig.~\ref{fig:muf}, the $C^{\text{OM}}\otimes f_{u-d}$ has small non-zero values outside the region $-1<x<1$.  To examine these differences more closely, we subtract the central curve $xf_{u-d}(x,\mu=3\text{ GeV})$ from both $xf_{u-d}(x,\mu)$ and $x(C^{\text{OM}}\otimes f_{u-d})$, and plot their differences in the right panel of Fig.~\ref{fig:muf}. 
 
\begin{figure*}[t!]
	\centering
	\includegraphics[width=0.49\textwidth]{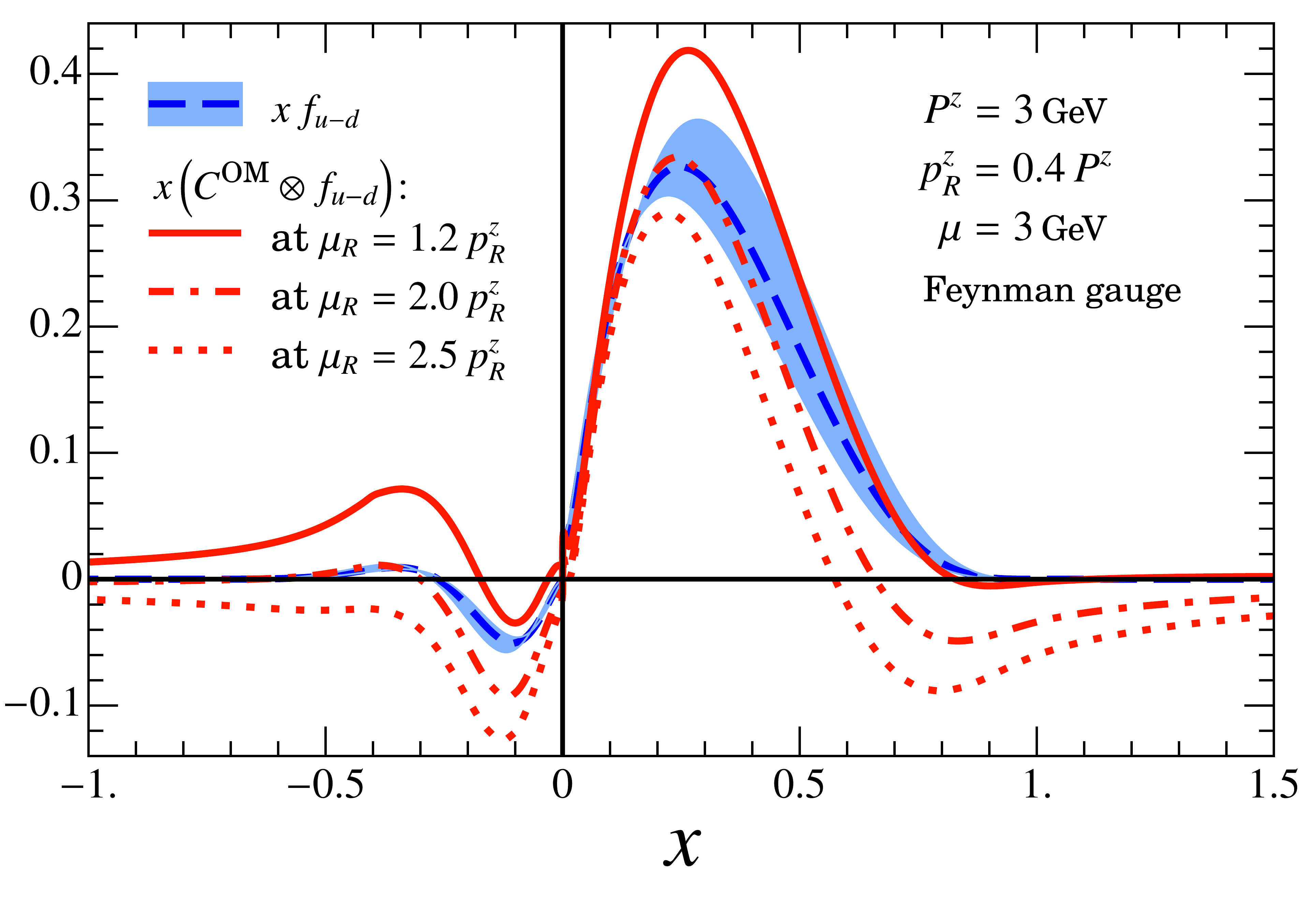}
	\hspace{0.1cm}
 	\includegraphics[width=0.49\textwidth]{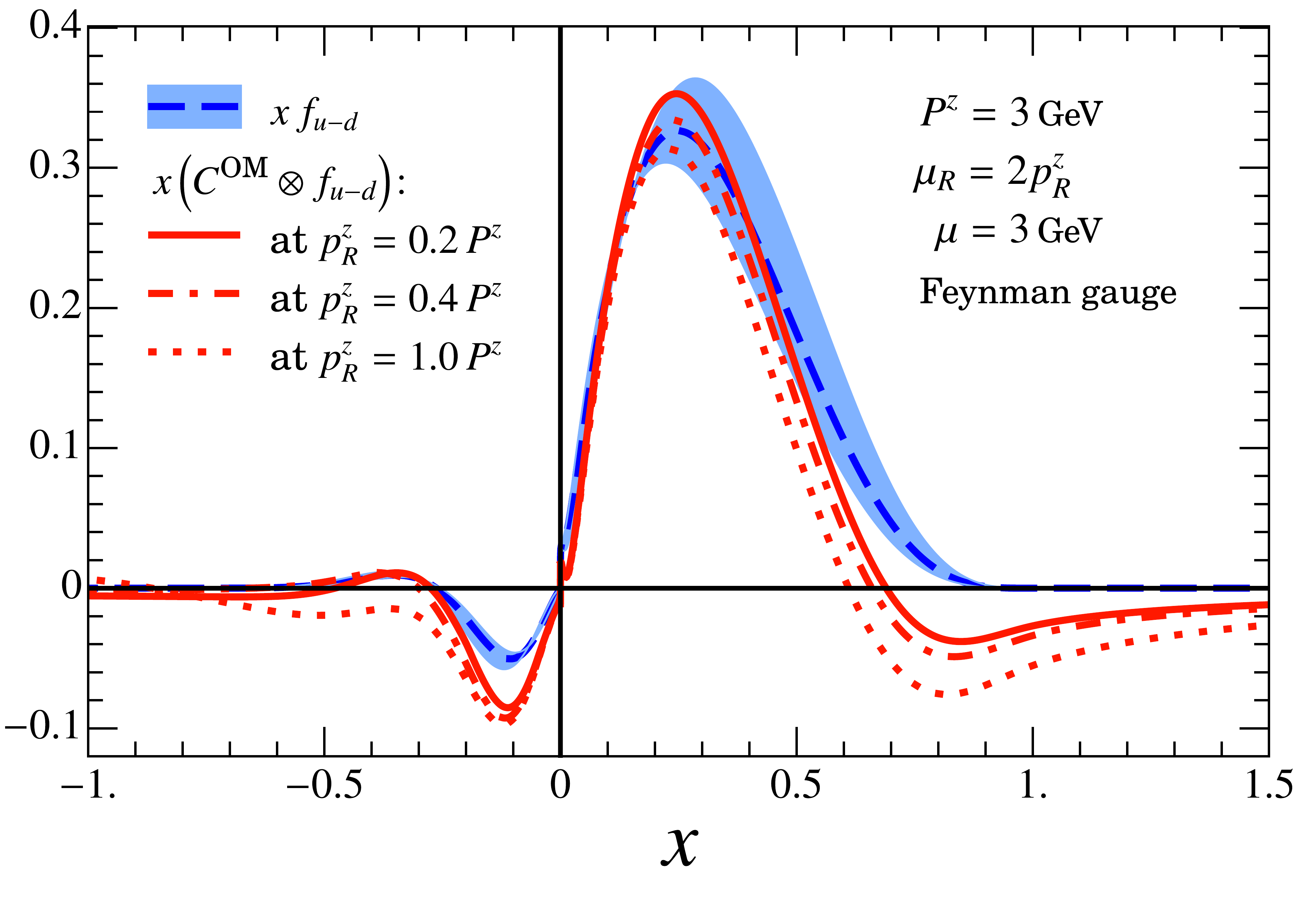}
	\caption{Left panel: Comparison between the PDF $xf_{u-d}$ and the quasi-PDF from $x(C^{\text{OM}}\otimes f_{u-d})$ determined at different $\mu_R$s.  Right panel: The $p_R^z$ dependence of the quasi-PDF $x(C^{\text{OM}}\otimes f_{u-d})$, compared to the PDF $xf_{u-d}$ which is independent of $p_R^z$. In both panels the blue band indicates the $\mu$ renormalization scale dependence of the PDF from variation by a factor of two.}
	\label{fig:mur}
\end{figure*}

\begin{figure}[t!]
	\centering
	\includegraphics[width=0.49\textwidth]{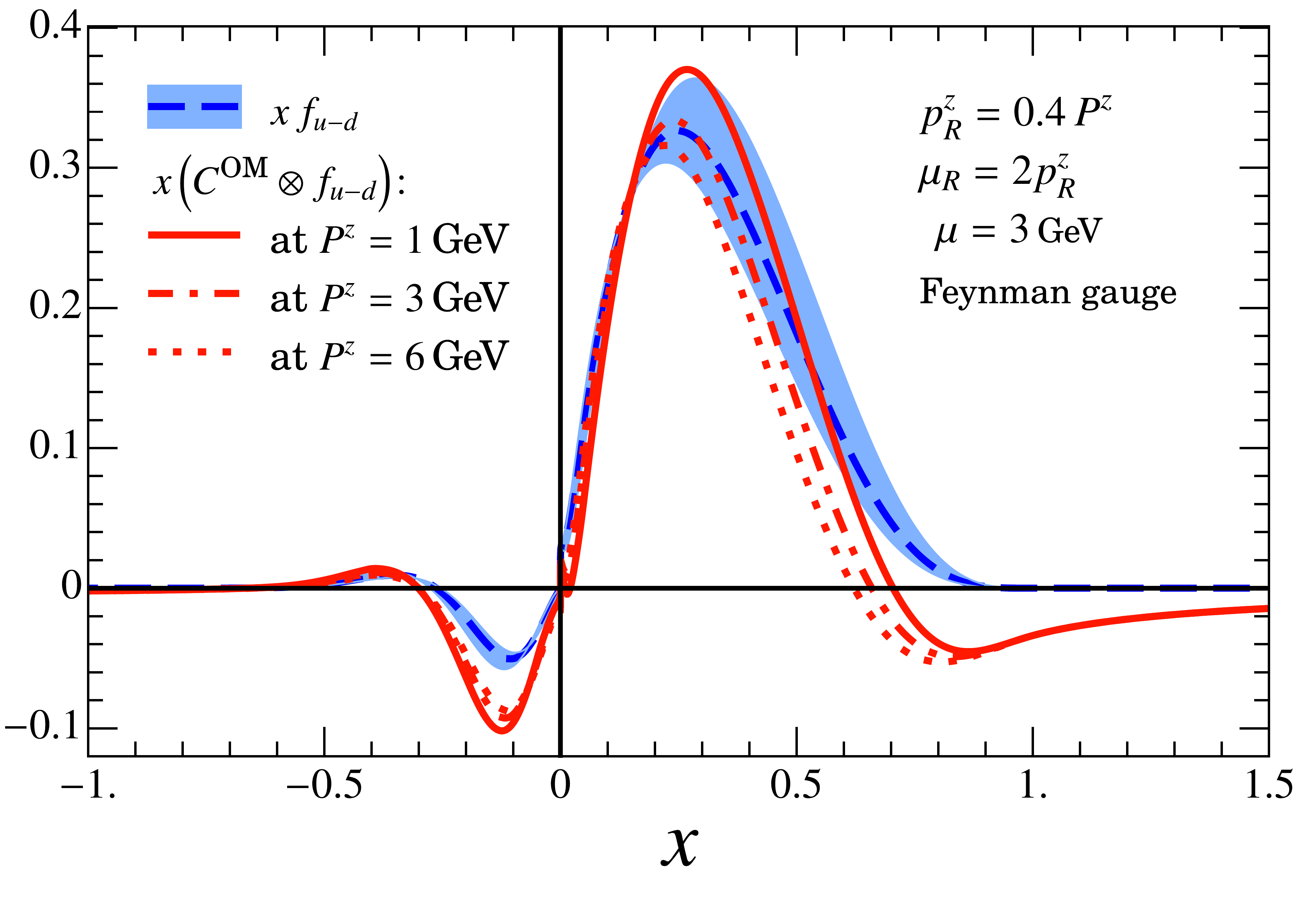}
	\caption{Comparison between the PDF $xf_{u-d}$ and the quasi-PDF from $x(C^{\text{OM}}\otimes f_{u-d})$ determined at different $P^z$s. The blue band indicates the $\mu$ renormalization scale dependence of the PDF from variation by a factor of two.}
	\label{fig:Pz}
\end{figure}

Next we examine the gauge dependence of the quasi-PDF in the RI/MOM scheme. The result for Landau gauge ($\tau=0$) is shown in the left panel of Fig.~\ref{fig:landaugauge}, which is plotted in the same way as Fig.~\ref{fig:muf}, and appears very similar.  To examine the change to the quasi-PDF we therefore plot the difference between the Landau gauge and Feynman gauge result in the right panel of Fig.~\ref{fig:landaugauge}. We see that
the $\tau$-dependent contribution in the matching coefficient is a fairly small but noticeable correction in the $0<x<1$ region, but can be a larger correction for negative $x$. 

Our next step is to fix the factorization scale at $\mu=3.0$ GeV and vary the parameters $r_R$ and $p_R^z$ in the quasi-PDF. From Eq.~(\ref{renqpdf}) we see that the definition of the RI/MOM quasi-PDF depends on the parameter $r_R$, so different $r_R$s and $p_R^z$s correspond to different quasi-PDfs. For $\mu_R=\{1.2, 2.0, 2.5\}p_R^z$, we plot in the left panel of Fig.~\ref{fig:mur} a comparison between $x(C^{\text{OM}}\otimes f_{u-d})(x,\mu_R)$ and $xf_{u-d}(x,\mu)$ with $\mu=3.0\,{\rm GeV}$ (blue dashed line). The blue band shows how the PDF changes when we vary $\mu$ between $\mu=1.5$ and $\mu=6.0\,{\rm GeV}$. We see that the RI/MOM quasi-PDF is quite sensitive to the choice of $r_R$, exhibiting larger variations than that of varying the renormalization scale $\mu$ in the PDF.  We also observe that the quasi-PDF moves away from the PDF as $\mu_R/p_R^z$ is made larger. The quasi-PDF in the RI/MOM scheme also satisfies a multiplicative renormalization group equation, derived in \app{rge}, which can be analyzed with a perturbative anomalous dimension for $\mu_R\gg \Lambda_{\rm QCD}$.  In the right panel of Fig.~\ref{fig:mur} we vary $p^z_R = \{0.2,0.4,1.0\} P^z$ while holding fixed $\mu_R=2.0 p_R^z$ and $P^z=\mu=3.0$ GeV. We observe that there exists a
range of values with $p_R^z/P^z \simeq 0.2-0.4$ which tend to minimize the impact of the matching coefficient in the $0<x<1$ region.

In Fig.~\ref{fig:Pz} we vary $P^z=\{1, 3, 6\}$ GeV while holding $\mu_R=2.0p_R^z$, $p_R^z=0.4P^z$ fixed with $\mu=3.0$ GeV. We observe that in the tails ($x>1$ and $x<-1$) that the RI/MOM quasi-PDF is not sensitive to $P^z$ . On the other hand, in the central region $-1<x<1$ the matching coefficient gives non-trivial corrections in the $P^z\to \infty$ limit, and hence there is always perturbative conversion needed between the quasi-PDF and PDF.

\begin{figure}[t!]
\centering
\includegraphics[width=0.49\textwidth]{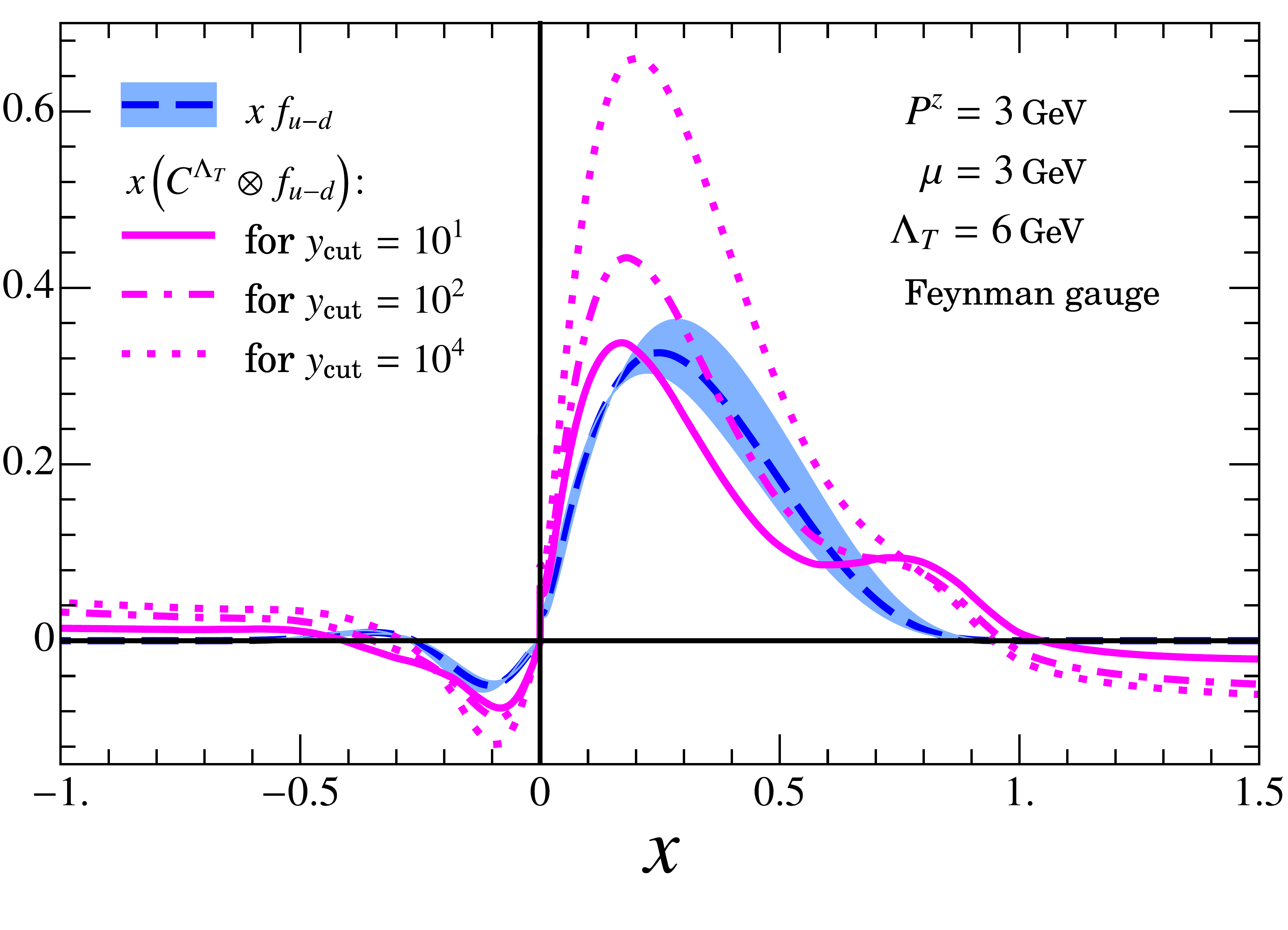}
\caption{Results for the quasi-PDF in the transverse cutoff scheme compared to the PDF $f_{u-d}$.  We show $x(C^{\Lambda_T}\otimes f_{u-d})$ with three different values for the cutoff $y_\text{cut}$ in the convolution integral.}
\label{fig:otherschemes}
\end{figure}

Finally we consider the comparison between our RI/MOM results and the matching results in the transverse cut-off scheme. At $P^z=\mu=3.0$ GeV, $p_R^z=0.4P^z$, $\mu_R=2.0p_R^z$, and $\Lambda_{\rm T}=6.0\,{\rm GeV}$, we calculate $C^{\Lambda_{\rm T}}\otimes f_{u-d}$ with $y_{\text{cut}}=10^1, 10^2, 10^4$ and plot the results with comparison to $f_{u-d}$ in Fig.~\ref{fig:otherschemes}. Unlike for $C^{\text{OM}}\otimes f_{u-d}$, the results for $C^{\Lambda_{\rm T}}\otimes f_{u-d}$  suffer from UV divergences in the integration over $y$, and they differ significantly from $f_{u-d}$. This means that when one inverts the factorization formula in Eq.~(\ref{eq:factorization}) to determine the PDF from the quasi-PDF, that there must be a large cancellation of UV divergences between the quasi-PDF from lattice QCD and the matching coefficient in the convolution integral. Since the UV region of a matching factor $C(x/y)$ is near $y=0$ as well as $y\to\infty$ for its $\delta$-function part, it is necessary to test the sensitivity of the convolution integral to the smallest momentum fraction of the quasi-PDF and $y_{\rm cut}$ for the lattice calculations in Ref.~\cite{Lin:2014zya,Alexandrou:2015rja,Chen:2016utp,Alexandrou:2016jqi,Alexandrou:2017huk}. Using the RI/MOM scheme avoids this complication.  Another advantage of the RI/MOM scheme is that in the unphysical regions $|x| > 1$ the $\alpha_s$ corrections to the matching coefficient fall as $1/(P^z)^2$ as $P^z\to \infty$, so the quasi-PDF will vanish asymptotically in this region. This is not the case for the quasi-PDF in the transverse cutoff scheme, where the quasi-PDF asymptotes to a non-trivial function for $|x|>1$ when $P^z\to \infty$. 

The reason why $C^{\text{OM}}\otimes f_{u-d}$ has better UV convergence than the transverse cutoff scheme is that the RI/MOM scheme introduces a counterterm to the quasi-PDF which cancels out the UV divergences that arises when one integrates over $y$. The matching result in the RI/MOM scheme therefore yields only a bounded small effect which one can be more confident about treating perturbatively. Thus, to reduce the uncertainties, our results imply that it is reasonable to favor RI/MOM over the transverse cutoff scheme. (In Ref.~\cite{Izubuchi:2018srq} it is shown that the matching coefficient for the quasi-PDF in the $\overline{\rm MS}$ scheme, $C^{\overline{\rm MS}}$, also yields a convergent convolution integral.)

\section{conclusion}
\label{sec:conclusion}

We have described the procedure of nonperturbative renormalization of quasi-PDF in the RI/MOM scheme. The $z$-dependent renormalization constant is obtained by imposing Eq.~(\ref{eq:rimom}) on the off-shell quark matrix element of the spatial correlation operator in lattice QCD. Then the renormalization constant is applied to the nucleon matrix element of the same correlation operator whose Fourier transform gives the quasi-PDF on the lattice. In RI/MOM the renormalized quasi-PDF is regularization invariant and can be related to PDF in the $\overline{\text{MS}}$ scheme through a perturbative matching condition, which is calculable in the continuum theory with dimensional regularization. Since all the large corrections in lattice perturbation theory are absorbed into the nonperturbative renormalization constant, the uncertainty of this procedure comes from lattice discretization effects and perturbative matching in the continuum theory. Our numerical results show that the one-loop matching for the RI/MOM scheme has nice UV convergence and reasonable magnitude for a perturbative correction, which is in contrast to the matching result for the transverse-momentum cutoff scheme. This indicates that the theoretical uncertainty in the perturbative matching for the RI/MOM scheme is small and controllable, thus making it more favorable than the transverse-momentum cutoff scheme. Furthermore, the matching in the RI/MOM scheme is consistent with the quasi-PDF vanishing in the unphysical region $|x|>1$ as $P^z\to \infty$, unlike the results in the transverse cutoff scheme. To increase the accuracy of our results in the future, one can study the $O(a)$ improvement for the lattice simulation of quasi-PDF and calculate the matching coefficient to higher orders in perturbation theory. A crucial ingredient in obtaining more accurate lattice results is to consider larger $P^z$ by working with finer lattices, in order to reduce power corrections. As the capabilities for doing simulations with larger nucleon momentum on the lattice continue to improve, we believe that our results will provide an important ingredient for future lattice calculations of PDF with the desired accuracy.

It should be noted that on the lattice there are not only discretization errors, but also mixings between the quasi-PDF and other non-local operators due to the broken symmetries. We should include all the possible operators that mix with the gauge-invariant quark bilinear and determine their renormalization constants nonperturbatively~\cite{Alexandrou:2017huk,Constantinou:2017sej,Chen:2017mzz,Green:2017xeu}. (Note that this is much simpler than the mixing between local operators considered in~\cite{Rossi:2017muf}, since there is no mixing to lower dimensional non-local operators. The mixing between local operators plays no role in the quasi-PDF analysis.)  We must also be aware of the fact that one can only calculate the spatial correlation for the quasi-PDF at a finite number of discrete $z$'s on the lattice. As a result, its Fourier transform into momentum space will exhibit an oscillatory behavior due to the truncation at $|z_\text{max}|$~\cite{Alexandrou:2017huk,Chen:2017mzz,Green:2017xeu}. 
Nevertheless, if we derive the parametric behavior of the spatial correlation at large $|z|$, we can use a proper set of basis functions to fit the data points and obtain a smooth curve. Then we can Fourier transform this smooth curve to obtain the quasi-PDF, which is free of the truncation error and should capture the correct behavior in the small $x$ region that can only be probed at large $|z|$. The choice of the basis functions could be based on results from global fits to the PDF with matching correction in Eq.~(\ref{eq:factorization}). Two other approaches to reduce the truncation error were also proposed in the recent paper~\cite{Lin:2017ani}.

\section*{Acknowledgement}
The authors are thankful for discussions with J.~W.~Chen, W.~Detmold, M.~Constantinou, L.~Jin, Y.~B.~Yang, K.~F.~Liu, J.~H.~Zhang, M.~Engelhardt, P.~Shanahan and D. Horkel. This material was supported by the U.S. Department of Energy, Office of Science, Office of Nuclear Physics, from DE-SC0011090 and within the framework of the TMD Topical Collaboration.  I.S. was also supported in part by the Simons Foundation through the Investigator grant 327942.

\appendix

\section{Vector Current Conservation}
\label{sec:conservation} 

In dimensional regularization and the Feynman gauge, the one-loop correction to the local vector current in an off-shell quark state $|p\rangle$ is
\begin{align}\label{eq:vec}
\delta \Gamma^z = {\alpha_s C_F\over 4\pi}\left[ \left({1\over\epsilon} + \ln{\mu^2\over -p^2}+1\right)\gamma^z - {2p^z\slashed p\over p^2} \right] + \delta Z_\psi \gamma^z \ .
\end{align}
The quark self-energy graph gives
\begin{align}
\Sigma(p) = -{\alpha_s C_F\over 4\pi} \left[ {1\over\epsilon} + \ln{\mu^2\over -p^2} +1 \right] \slashed p  = \delta Z_\psi\, \slashed p \, ,
\end{align}
where $\delta Z_\psi$ is defined by the correction to the residue of the pole in the propagator. Thus for the on-shell renormalization of the wavefunction in Eq.~(\ref{onshell}), we find that the $\gamma^z$ terms exactly cancel in $\delta\Gamma^z$, but the $\slashed p$ term in Eq.~(\ref{eq:vec}) remains. The correction to the vector current $\bar{\psi}\gamma^z\psi$ is zero, but there is mixing to the scalar operator $\bar{\psi}\slashed p \psi$.

If we define the off-shell matrix element using \eq{pslash}, then 
\begin{align}
\delta \Gamma'^z = {\alpha_s C_F\over 4\pi}\left[ \left({1\over\epsilon} + \ln{\mu^2\over -p^2}+1\right) - 2 \right] + \delta Z_\psi \ ,
\end{align}
where the $\slashed p$ term has made a nonzero contribution. Nevertheless, if we redefine $\delta Z_\psi$ according to Eq.~(\ref{ward}), 
\begin{align}
\delta Z'_\psi = -{\alpha_s C_F\over 4\pi}\left[ \left({1\over\epsilon} + \ln{\mu^2\over -p^2}+1\right) - 2 \right] \,,
\end{align}
then the one-loop corrections to the vector current exactly cancel. Actually, this cancellation takes place even before loop integration as it is guaranteed by the Ward-Takahashi identity.

\section{Matching On-shell}
\label{sec:on-shell}

The operator mixing, non-vanishing one-loop correction, and gauge-dependence are all due to the off-shellness of the external state. In this appendix we show that the same result is obtained for the matching coefficient $C$ when we use dimensional regularization as the IR regulator.

If we work on-shell, using dimensional regularization for both UV and IR divergences, then we can use the equation of motion to eliminate the $\slashed p$ term and there will be no difference in the two definitions of $\delta Z_{\psi}$ in Eqs.~(\ref{onshell},\ref{ward}). With a massless quark the vertex renormalization for the local vector current is
\begin{align}
\delta \Gamma^z = {\alpha_s C_F\over 4\pi}\left[ \left({1\over\epsilon_{\rm UV}} - {1\over \epsilon_{\rm IR}}\right)\gamma^z \right] + \delta Z_\psi \gamma^z \ ,
\end{align}
and
\beq
\Sigma(p) = -{\alpha_s C_F\over 4\pi} \left[ {1\over\epsilon_{\rm UV}} -{1\over \epsilon_{\rm IR}}\right] \slashed p\ ,
\eeq
where $1/\epsilon_{\rm UV}$ and $1/\epsilon_{\rm IR}$ denote the UV and IR divergences respectively. The two definitions in Eqs.~(\ref{onshell},\ref{ward}) give the same $\delta Z_\psi$, and one-loop correction to the vector current is strictly guaranteed to be zero. However, the off-shell momentum subtraction scheme cannot be implemented on-shell.  Nevertheless, the matching calculation requires comparing renormalized matrix elements for the quasi-PDF and PDF, so we can carry out the matching on-shell with the $1/\epsilon_{\rm IR}$ IR regulator, where the renormalized quasi-PDF is defined using the UV counterterm from the off-shell momentum subtraction scheme.

For on-shell massless quarks, the unrenormalized quasi-PDF calculated in this fashion is
\begin{widetext}
\begin{align} \label{eq:qtonshellbare}
\tilde{q}^{(1)}_\epsilon(x,p^z)&= {\alpha_s C_F\over 2\pi}(4\zeta)\left\{\begin{array}{ll}
\displaystyle \bigg[ {1+x^2\over 1-x}\ln{x\over x-1} +1 \bigg]_\oplus \ 
 & x>1\\[12pt]
\displaystyle \bigg[ {1+x^2\over 1-x}\biggl(-{1\over \epsilon_{\rm IR}} - \ln\frac{\mu^2}{p_z^2}+\ln[4x(1-x)]\biggr) +  2-x- {2x\over 1-x} \bigg]_+ \ 
 & 0<x<1 \\[12pt]
\displaystyle\bigg[ -{1+x^2\over 1-x}\ln{x\over x-1} -1 \bigg]_\ominus   \  & x<0
\end{array}\right.  .
\end{align}

By adding the counter-term defined in Eq.~(\ref{counterterm}), we have the renormalized quasi-PDF
\begin{align}
&\tilde{q}^{(1)}_\epsilon\Big(x,p^z,p_R^z,\mu_R,\mu\Big)
={\alpha_s C_F\over 2\pi}(4\zeta)\nonumber\\
&\times\!\left\{\begin{array}{ll}
\displaystyle \bigg[ {1+x^2\over 1-x}\ln{x\over x-1} - {2\over \sqrt{r_R-1}}\left[{1+x^2\over 1-x} -{r_R\over2(1-x)}\right]\arctan {\sqrt{r_R-1}\over 2x-1} +{r_R\over 4x(x-1)+r_R} \bigg]_\oplus \
 & x>1\\[12pt]
\displaystyle \bigg[ {1+x^2\over 1-x}\left(-{1\over \epsilon_{\rm IR}} - \ln\frac{\mu^2}{p_z^2}+\ln[4x(1-x)]\right)  + (2-x)
 -{2(1+x^2 -r_R/2)\over \sqrt{r_R-1}(1-x)} \arctan\sqrt{r_R-1}\bigg]_+ \ \
 & 0<x<1 \\[12pt]
\displaystyle\bigg[  -{1+x^2\over 1-x}\ln{x\over x-1}+ {2\over \sqrt{r_R-1}}\left[{1+x^2\over 1-x} -{r_R\over2(1-x)}\right]\arctan {\sqrt{r_R-1}\over 2x-1}- {r_R\over 4x(x-1)+r_R} \bigg]_\ominus \  & x<0
\end{array}\right. 
\nonumber \\
& \ \ + {\alpha_sC_F\over2\pi} (4\zeta) \bigg\{ 
  h(x,r_R) - |\eta|\, h\big(1 + \eta(x-1),r_R\big)
   \biggr\}
.
\end{align}
In pure dimensional regularization the bare PDF at one-loop is proportional to $(1/\epsilon_{\rm UV}-1/\epsilon_{\rm IR})$, so on-shell with dimensional regularization regulating the IR divergence the renormalized PDF in the $\overline{\text{MS}}$ scheme is
\begin{align} \label{eq:qonshellren}
q^{(1)}_\epsilon(x,\mu)=& {\alpha_s C_F\over 2\pi}(4\zeta)\left\{\begin{array}{ll}
\displaystyle0 \  & x>1\\
\displaystyle{ \left( {1+x^2 \over 1-x} \right)_+  \left(-{1\over \epsilon_{\rm IR}} \right)  } \ \ \
 & 0<x<1 \\
\displaystyle0 \  & x<0
\end{array}\right. \ .
\end{align}
Computing the matching coefficient
\beq
C^{\rm OM}\left(\xi,{\mu_R\over p_R^z},{\mu\over p^z},{p^z\over p_R^z}\right) = \delta(1-\xi) +{1\over 4\zeta}\left[\tilde{q}^{(1)}_\epsilon\left(\xi,p^z,p_R^z,\mu_R,\mu\right) - q^{(1)}_\epsilon\left(\xi,\mu\right)\right]
\eeq
then gives exactly the same result for $C^{\text{OM}}$ as in Eq.~(\ref{eq:crimom}). This gives an explicit example showing that the matching result is independent of the common IR regulator we choose to use for the PDF and quasi-PDF in the calculation.

\section{An Alternate RI/MOM Scheme}  \label{sec:altRIMOM}

Now let us turn to the alternate scheme for defining the RI/MOM quasi-PDF given in \eq{gammaz}. With this choice the unrenormalized one-loop matrix element in a general covariant gauge is given by our earlier result in \eq{covrt} plus an additional term
\begin{align}
\tilde{q}^{(1)}_{\gamma^z\tau}\bigl(x,p^z,0,p^2\bigr)
  &=\tilde{q}^{(1)}_{\tau}\bigl(x,p^z,0,p^2\bigr)
 + \Delta\tilde{q}_{\gamma^z\tau}^{(1)} \bigl(x,p^z,0,p^2\bigr)
  \,,
\end{align}
with
\begin{align} \label{eq:Delqgamz}
\Delta\tilde{q}^{(1)}_{\gamma^z\tau}\bigl(x,p^z,0,p^2\bigr)
 &={\alpha_s C_F\over 4\pi} (4\zeta)
 \left\{
\begin{array}{ll}
\displaystyle \Bigl[  \Delta h_{\gamma^z\tau}(x,\rho) \Bigr]_\oplus 
  \ \ \ 
  & x>1\\[15pt]
\displaystyle \Bigl[ \Delta h_{\gamma^z\tau}(x,\rho) \Bigr]_+
  &0<x<1\\[12pt]
\displaystyle \Bigl[  \Delta h_{\gamma^z\tau}(x,\rho) \Bigr]_\ominus
 & x<0
\end{array}\right. 
,
\end{align}
where
\begin{align}
  \Delta h_{\gamma^z\tau}(x,\rho) &\equiv
\left\{
\begin{array}{ll}
\displaystyle {\rho\over(1-\rho)^{3\over2}}{2x^2-\rho\over 1-x}\ln {2x-1 +\sqrt{1-\rho}\over 2x-1 - \sqrt{1-\rho}} + {2\rho\over 1-\rho} {\rho + 2x(x-1)(2x+1) \over (x-1)[\rho+4x(x-1)]}
 +{2(1-\tau)\rho^2\over \bigl[\rho+4x(x-1)\bigr]^2}
  & x>1\\[15pt]
\displaystyle {\rho\over(1-\rho)^{3\over2}}{2x^2-\rho\over 1-x}
  \ln {1+\sqrt{1-\rho}\over 1- \sqrt{1-\rho}} 
 + {2\over 1-\rho}{\rho(1-2x)+2x(1-x)\over 1-x} 
 - 2 (1-\tau) 
  &\!\!\!\!\!\! 0<x<1\\[12pt]
\displaystyle -{\rho\over(1-\rho)^{3\over2}}{2x^2-\rho\over 1-x}\ln {2x-1 +\sqrt{1-\rho}\over 2x-1 - \sqrt{1-\rho}} - {2\rho\over 1-\rho} {\rho + 2x(x-1)(2x+1) \over (x-1)[\rho+4x(x-1)]} 
 -{2(1-\tau)\rho^2\over \bigl[\rho+4x(x-1)\bigr]^2} \ \ 
 & x<0
\end{array}\right. .
\end{align}

In the RI/MOM scheme, combining the matrix element with $\rho\to 0$ with the counterterm, the correction to the renormalized matrix element is
\begin{align}
\Delta\tilde{q}^{(1)}_{\gamma^z\tau}\bigl(x,p^z,p_R^z,\mu_R\bigr)
 &= - {\alpha_s C_F\over 4\pi} (4\zeta)
 \left\{
\begin{array}{ll}
\displaystyle \Bigl[  \Delta h_{\gamma^z\tau}(x,r_R) \Bigr]_\oplus 
  \ \ \ 
  & x>1\\[15pt]
\displaystyle \Bigl[ \Delta h_{\gamma^z\tau}(x,r_R) +2(1-\tau)-4x \Bigr]_+
  &0<x<1\\[12pt]
\displaystyle \Bigl[  \Delta h_{\gamma^z\tau}(x,r_R) \Bigr]_\ominus
 & x<0
\end{array}\right. \!\! 
 \nn\\
&\ \ + {\alpha_s C_F\over 4\pi} (4\zeta)
 \Bigl[ \Delta h_{\gamma^z\tau}(x,r_R) - |\eta|\, \Delta h_{\gamma^z\tau}\bigl(1+\eta(x-1),r_R \bigr)
 \Bigr]
. 
\end{align}
To carry out the matching we must implement the same infrared regulator (ie. treatment of spinors in the off-shell regulator) for the PDF calculation. To do this we decompose the light-cone PDF into $\{\gamma^z, \slashed{p}\}$ terms, and only keep the $\gamma^z$ terms. Since $\gamma^z = (\gamma^+ - \gamma^-)/2$, and $\bar u(p) \gamma^- u(p) \to 0$ in a linear fashion as $p^2\to 0$, the $\gamma^z$ coefficient turns out to be equivalent to the coefficient of $\gamma^+$ with a basis of $\{\gamma^+, \slashed{p}\}$. 
Extracting the coefficient of $\gamma^+$ in this basis, the PDF matrix element calculation is modified as\footnote{We note that our calculation of the PDF with an off-shell regulator also agrees with the calculation of this quantity in Ref.~\cite{Stewart:2010qs}. To achieve this we decompose the PDF in a $\{\gamma^+, \gamma^-\}$ basis and then define the PDF as the $\gamma^+$ coefficient, while also accounting for the different choice for the wavefunction renormalization in Eqs.~(\ref{onshell}) versus (\ref{ward}).} 
\begin{align}
q_{\gamma^+\tau}^{(1)}(x,\mu)&=q^{(1)}(x,\mu)+{\alpha_sC_F\over 4\pi} (4\zeta)
\left\{\begin{array}{cl}
\displaystyle 0\ & x>1\\
\displaystyle \left[4x - 2(1-\tau)\right]_+\ \ &0<x<1\\
\displaystyle 0\ & x<0
\end{array}\right.\ .
\end{align}
\begin{figure*}
	\centering
	\includegraphics[width=0.49\textwidth]{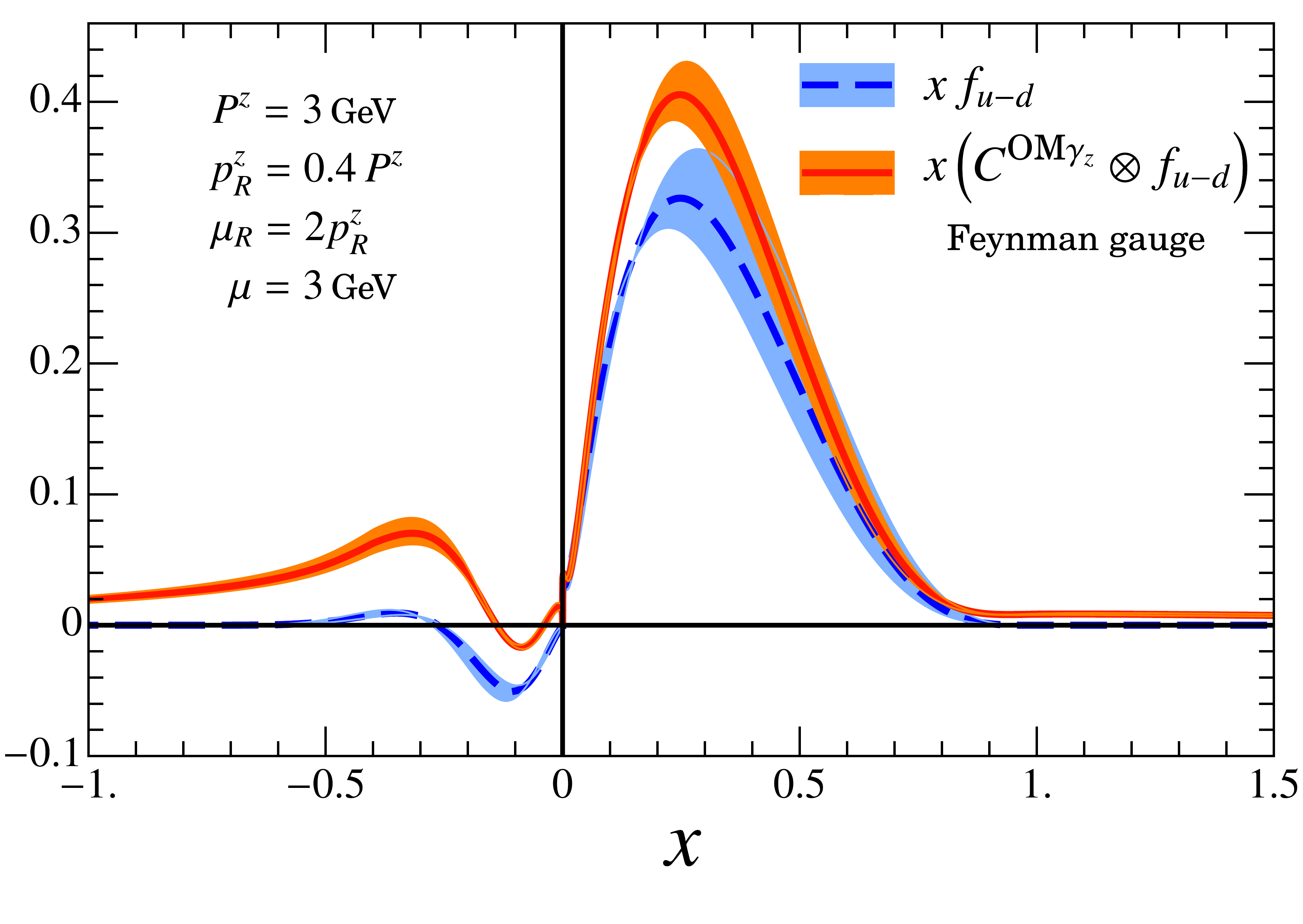}
	\hspace{0.1cm}
	\includegraphics[width=0.49\textwidth]{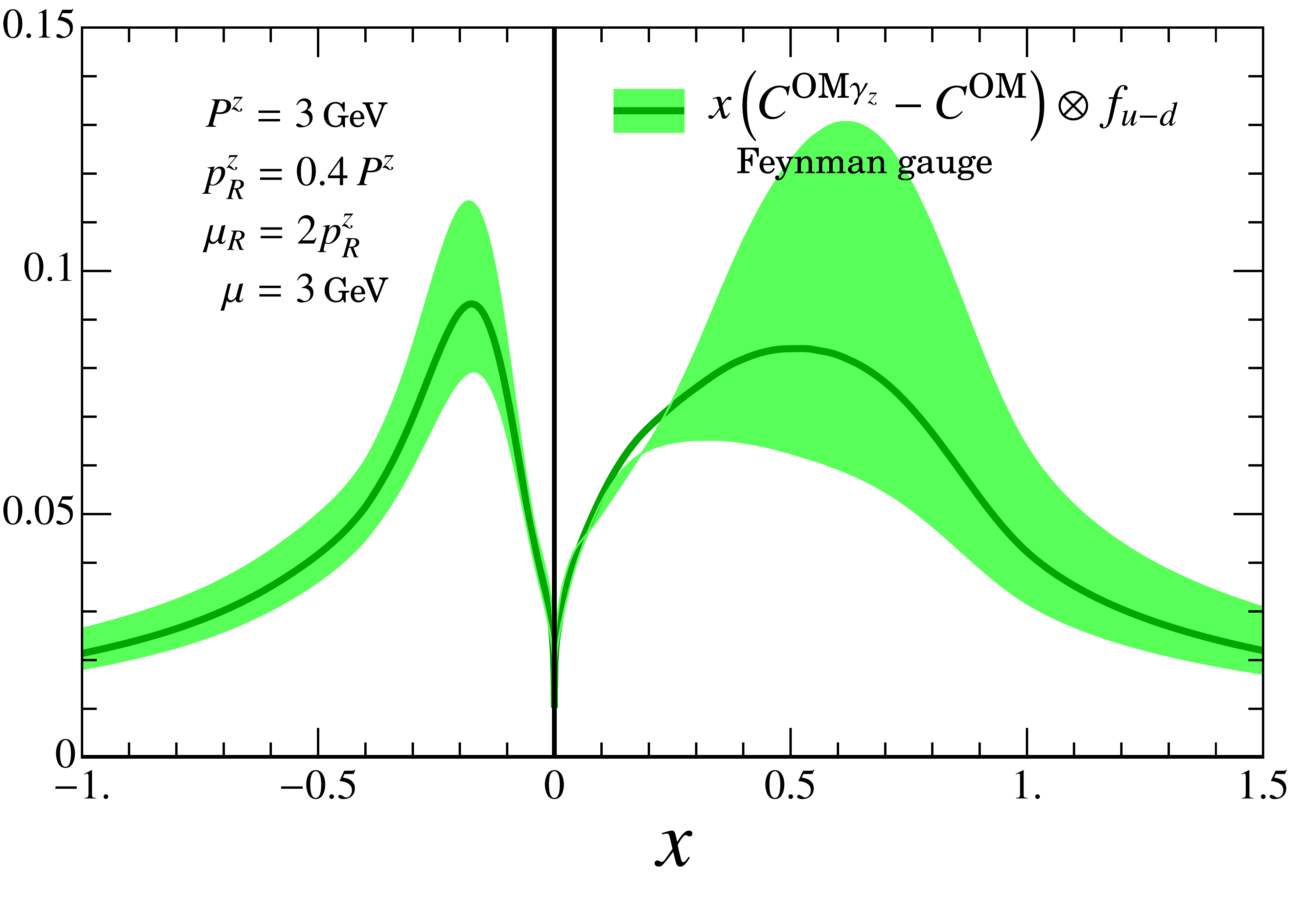}
	\caption{Comparison between the PDF $xf_{u-d}$ and the quasi-PDF obtained from $x(C^{\text{OM}\gamma^z}\otimes f_{u-d})$ in the Feynman gauge. The yellow, blue and green bands indicate the uncertainties in the factorization scale $\mu$. Left: $x(C^{\text{OM}\gamma^z}\otimes f_{u-d})$ and $xf_{u-d}$. Right: $x(C^{\text{OM}\gamma^z}-C^{\text{OM}})\otimes f_{u-d}$.}
	\label{fig:gammaz}
\end{figure*}
Using these results the matching coefficient becomes
\beq
C^{\text{OM}\gamma^z}\left(\xi, {\mu_R\over p_R^z},{\mu\over p^z},{p^z \over p_R^z}\right) = C^{\text{OM}}\left(\xi, {\mu_R\over p_R^z},{\mu\over p^z},{p^z \over p_R^z}\right) + \Delta C^{\text{OM}\gamma^z}\left(\xi, {\mu_R\over p_R^z},{p^z \over p_R^z}\right)
\eeq
with 
\begin{align} \label{eq:DelCgamz}
&\Delta C^{\text{OM}\gamma^z}\left(\xi, {\mu_R\over p_R^z},{p^z \over p_R^z}\right)
 ={\alpha_s C_F\over 4\pi}
 \nn\\
&\times \left\{
\begin{array}{ll}
\displaystyle \left[{2r_R\over(r_R-1)^{3\over2}}{2\xi^2-r_R\over (1-\xi)}\arctan{\sqrt{r_R-1}\over 2\xi-1}+{2r_R\over r_R-1} {r_R + 2\xi(\xi-1)(2\xi+1) \over (\xi-1)[r_R+4\xi(\xi-1)]}- {2(1-\tau) r_R^2 \over \bigl[r_R+4x(x-1)\bigr]^2}\right]_\oplus 
& \xi>1\\
\displaystyle \left[{2r_R\over(r_R-1)^{3\over2}}{2\xi^2-r_R\over (1-\xi)}\arctan\sqrt{r_R-1} + {2\over r_R-1}{r_R(1-2\xi)+2\xi(1-\xi)\over 1-\xi}+2(1-\tau)\right]_+ 
&0<\xi<1\\
\displaystyle \left[-{2r_R\over(r_R-1)^{3\over2}}{2\xi^2-r_R\over (1-\xi)}\arctan{\sqrt{r_R-1}\over 2\xi-1}-{2r_R\over r_R-1} {r_R + 2\xi(\xi-1)(2\xi+1) \over (\xi-1)[r_R+4\xi(\xi-1)]}
 + {2(1-\tau) r_R^2\over \bigl[r_R+4x(x-1) \bigr]^2 }  \right]_\ominus
  & \xi<0
\end{array}\right. 
 \nn\\
&\ \ + {\alpha_s C_F\over 4\pi} 
 \Bigl[ \Delta h_{\gamma^z\tau}(x,r_R) - |\eta|\, \Delta h_{\gamma^z\tau}\bigl(1+\eta(x-1),r_R \bigr)
 \Bigr]
.
\end{align}
Again this result is independent of the choice of IR regulator used for the matching calculation. If we repeat this calculation using the on-shell scheme discussed in \app{on-shell} then it is easy to see that $\Delta C^{\text{OM}\gamma^z}(\xi,\mu_R/p_R^z) = [\tilde q^{(1)\gamma^z}_{\rm CT}(\xi,p_R^z,\mu_R)-\tilde q^{(1)}_{\rm CT}(\xi,p_R^z,\mu_R)]/(4\zeta)$, which is the difference of the counterterms that specify the two RI/MOM schemes.  Using \eq{Delqgamz} this immediately yields \eq{DelCgamz}.

Using the same settings as in Fig.~\ref{fig:muf}, we plot in the left panel of \fig{gammaz} the result for the quasi-PDF computed with the matching in this $\gamma^z$ scheme, $x(C^{\text{OM}\gamma^z}\otimes f_{u-d})$,  and compare it with the $\overline{\rm MS}$ PDF. In the right panel of  \fig{gammaz} we plot the difference between the two treatments of the spinors, $x(C^{\text{OM}\gamma_z}-C^{\text{OM}})\otimes f_{u-d}$. The modification to the matching from this alternate RI/MOM scheme is seen to be an effect of comparable size to the difference between the quasi-PDF and PDF shown in Fig.~\ref{fig:muf} (right panel), and hence it is important to carefully specify the treatment of spinors when defining an RI/MOM scheme. The results for the two schemes given here suffice to determine the result in any other possible scheme defined by a different treatment of the spinors at one-loop order.

\end{widetext}

\section{Renormalization Group Equation for the RI/MOM Quasi-PDF}
\label{sec:rge}

The $\mu_R$ dependence of the quasi-PDF takes a rather nontrivial form in the RI/MOM scheme, as shown in Eq.~(\ref{eq:renormzpdf}). It is useful to examine this dependence in the form of a renormalization group equation (RGE). Due to the multiplicative renormalization in Eq.~(\ref{eq:qPDFposZ}), the RGE of the quasi-PDF in position space has the form
\begin{align} \label{eq:rge}
	{d\tilde{q}(z,P^z,p_R^z,\mu_R)\over d\ln\mu_R}
   = \tilde{\gamma}(z,p_R^z, \mu_R)\: \tilde{q}(z,P^z,p_R^z,\mu_R)\,,
\end{align}
where the anomalous dimension $\tilde \gamma$ can be computed from the renormalization constant through
\begin{align} \label{eq:anmldim}
	\tilde{\gamma}(z,p_R^z, \mu_R) = - {d\over d\ln\mu_R}\, \ln \tilde{Z}^{\rm OM}(z,p_R^z,\epsilon, \mu_R) \,.
\end{align}
Based on the result in Eq.~(\ref{eq:renormzpdf1}), and the fact that at one-loop  
\begin{align}
\tilde{\gamma}^{(1)}(z,p_R^z,\mu_R) 
&=\mu_R\, \frac{d}{d\mu_R} \: \frac{\tilde{q}^{(1)}_{\rm{CT}}(z,p^z,p_R^z,\mu_R)}{(4p^z\zeta\, e^{-i p^z z})} \,,
\end{align} 
we find
\begin{widetext}
\begin{align}\label{eq:renge}
\tilde{\gamma}^{(1)}(z,p_R^z,\mu_R) 
&= \frac{\alpha_s C_F}{2\pi} \int_{-\infty}^\infty dx 
\left(e^{i(1-x) p_R^z z} - 1\right)
 \\
&\quad 
\times \left\{\begin{array}{ll}
\displaystyle \biggl[  {2x-1\over x-1}{r_R\over r_R-1} {2(1+x^2)-r_R\over r_R +4x(x-1)} + {8r_R x(x-1)\over \left[r_R+ 4x(x-1)\right]^2} & 
  \\[11pt]
\displaystyle \quad  - {r_R\over (r_R-1)^{3/2}}{2x^2+r_R\over x-1}\arctan{\sqrt{r_R-1}\over 2x-1}\, \biggr]\ 
 & x>1
  \\[15pt]
\displaystyle \biggl[ -{1\over r_R-1} {2(1+x^2)-r_R\over 1-x} + {r_R\over (r_R-1)^{3/2}}{2x^2+r_R\over 1-x}\arctan\sqrt{r_R-1}\, \biggr]\ \
 &0<x<1
  \\[15pt]
\displaystyle \biggl[  - {2x-1\over x-1}{r_R\over r_R-1} {2(1+x^2)-r_R\over r_R +4x(x-1)} - {8r_R x(x-1)\over \left[r_R+ 4x(x-1)\right]^2} & 
  \\[10pt]
\displaystyle \quad  + {r_R\over (r_R-1)^{3/2}}{2x^2+r_R\over x-1}\arctan{\sqrt{r_R-1}\over 2x-1}\, \biggr]\ 
 &x<0
\end{array}\right. \nn
\end{align}
where here $r_R = (\mu_R/p_R^z)^2$. 
Though hard to evaluate analytically, the above integral is finite and complex.

\end{widetext}

In the limit of $\mu_R\gg p_R^z$, we have
\begin{align} \label{eq:inftyrho}
\tilde{\gamma}^{(1)}(z,p_R^z,\mu_R) = & \frac{\alpha_s C_F}{4} \int\!\! {dx\over (1-x)} 
\left(e^{i(1-x)p_R^z z} - 1\right){\mu_R\over p_R^z} \,,
\end{align}
so the anomalous dimension does exhibit linear behavior in $\mu_R$. This shows that the RI/MOM scheme tracks the linear divergence from the Wilson line self-energy, and when solving \eq{rge} generates an exponential of $\mu_R/p_R^z$. (Equation~(\ref{eq:inftyrho}) appears to diverge for $|x|\gg \mu_R/p_R^z$, but this is just an artifact of expanding $ \mu_R\gg p_R^z$ with fixed $x$, and the true integrand in \eq{renge} goes to zero as $1/x^2$ for $x\to \pm \infty$.)

We can also consider the limit $\Lambda_{\rm{QCD}}\ll \mu_R \ll P^z$. In this limit the contributions from $x>1$ and $x<0$ vanish linearly, and analytically continuing to the region $r_R<1$, the result from the $0<x<1$ region is
\begin{align} \label{eq:smllrho}
&\tilde{\gamma}^{(1)}(z,P^z,\mu_R) \nn\\
&= \alpha_s C_F\int_0^1 {dx\over\pi} \left(e^{i(1-x)p_R^z z} - 1\right){1+x^2\over 1-x} \,,
\end{align}
which is just the Fourier transform of the standard DGLAP momentum space anomalous dimension.

\bibliography{quasiPDF}

\begin{thebibliography}{48}%
\makeatletter
\providecommand \@ifxundefined [1]{%
 \@ifx{#1\undefined}
}%
\providecommand \@ifnum [1]{%
 \ifnum #1\expandafter \@firstoftwo
 \else \expandafter \@secondoftwo
 \fi
}%
\providecommand \@ifx [1]{%
 \ifx #1\expandafter \@firstoftwo
 \else \expandafter \@secondoftwo
 \fi
}%
\providecommand \natexlab [1]{#1}%
\providecommand \enquote  [1]{``#1''}%
\providecommand \bibnamefont  [1]{#1}%
\providecommand \bibfnamefont [1]{#1}%
\providecommand \citenamefont [1]{#1}%
\providecommand \href@noop [0]{\@secondoftwo}%
\providecommand \href [0]{\begingroup \@sanitize@url \@href}%
\providecommand \@href[1]{\@@startlink{#1}\@@href}%
\providecommand \@@href[1]{\endgroup#1\@@endlink}%
\providecommand \@sanitize@url [0]{\catcode `\\12\catcode `\$12\catcode
  `\&12\catcode `\#12\catcode `\^12\catcode `\_12\catcode `\%12\relax}%
\providecommand \@@startlink[1]{}%
\providecommand \@@endlink[0]{}%
\providecommand \url  [0]{\begingroup\@sanitize@url \@url }%
\providecommand \@url [1]{\endgroup\@href {#1}{\urlprefix }}%
\providecommand \urlprefix  [0]{URL }%
\providecommand \Eprint [0]{\href }%
\providecommand \doibase [0]{http://dx.doi.org/}%
\providecommand \selectlanguage [0]{\@gobble}%
\providecommand \bibinfo  [0]{\@secondoftwo}%
\providecommand \bibfield  [0]{\@secondoftwo}%
\providecommand \translation [1]{[#1]}%
\providecommand \BibitemOpen [0]{}%
\providecommand \bibitemStop [0]{}%
\providecommand \bibitemNoStop [0]{.\EOS\space}%
\providecommand \EOS [0]{\spacefactor3000\relax}%
\providecommand \BibitemShut  [1]{\csname bibitem#1\endcsname}%
\let\auto@bib@innerbib\@empty
\bibitem [{\citenamefont {Collins}\ \emph {et~al.}(1988)\citenamefont
  {Collins}, \citenamefont {Soper},\ and\ \citenamefont
  {Sterman}}]{Collins:1989gx}%
  \BibitemOpen
  \bibfield  {author} {\bibinfo {author} {\bibfnamefont {J.~C.}\ \bibnamefont
  {Collins}}, \bibinfo {author} {\bibfnamefont {D.~E.}\ \bibnamefont {Soper}},
  \ and\ \bibinfo {author} {\bibfnamefont {G.}~\bibnamefont {Sterman}},\ }\href
  {http://arXiv.org/abs/hep-ph/0409313} {\bibfield  {journal} {\bibinfo
  {journal} {Adv. Ser. Direct. High Energy Phys.}\ }\textbf {\bibinfo {volume}
  {5}},\ \bibinfo {pages} {1} (\bibinfo {year} {1988})},\ \Eprint
  {http://arxiv.org/abs/hep-ph/0409313} {hep-ph/0409313} \BibitemShut {NoStop}%
\bibitem [{\citenamefont {Detmold}\ \emph {et~al.}(2001)\citenamefont
  {Detmold}, \citenamefont {Melnitchouk},\ and\ \citenamefont
  {Thomas}}]{Detmold:2001dv}%
  \BibitemOpen
  \bibfield  {author} {\bibinfo {author} {\bibfnamefont {W.}~\bibnamefont
  {Detmold}}, \bibinfo {author} {\bibfnamefont {W.}~\bibnamefont
  {Melnitchouk}}, \ and\ \bibinfo {author} {\bibfnamefont {A.~W.}\ \bibnamefont
  {Thomas}},\ }\href {\doibase 10.1007/s1010501c0013} {\bibfield  {journal}
  {\bibinfo  {journal} {Eur. Phys. J. direct}\ }\textbf {\bibinfo {volume}
  {3}},\ \bibinfo {pages} {1} (\bibinfo {year} {2001})},\ \Eprint
  {http://arxiv.org/abs/hep-lat/0108002} {arXiv:hep-lat/0108002 [hep-lat]}
  \BibitemShut {NoStop}%
\bibitem [{\citenamefont {Detmold}\ \emph {et~al.}(2002)\citenamefont
  {Detmold}, \citenamefont {Melnitchouk},\ and\ \citenamefont
  {Thomas}}]{Detmold:2002nf}%
  \BibitemOpen
  \bibfield  {author} {\bibinfo {author} {\bibfnamefont {W.}~\bibnamefont
  {Detmold}}, \bibinfo {author} {\bibfnamefont {W.}~\bibnamefont
  {Melnitchouk}}, \ and\ \bibinfo {author} {\bibfnamefont {A.~W.}\ \bibnamefont
  {Thomas}},\ }\href {\doibase 10.1103/PhysRevD.66.054501} {\bibfield
  {journal} {\bibinfo  {journal} {Phys. Rev.}\ }\textbf {\bibinfo {volume}
  {D66}},\ \bibinfo {pages} {054501} (\bibinfo {year} {2002})},\ \Eprint
  {http://arxiv.org/abs/hep-lat/0206001} {arXiv:hep-lat/0206001 [hep-lat]}
  \BibitemShut {NoStop}%
\bibitem [{\citenamefont {Dolgov}\ \emph {et~al.}(2002)\citenamefont {Dolgov}
  \emph {et~al.}}]{Dolgov:2002zm}%
  \BibitemOpen
  \bibfield  {author} {\bibinfo {author} {\bibfnamefont {D.}~\bibnamefont
  {Dolgov}} \emph {et~al.} (\bibinfo {collaboration} {LHPC, TXL}),\ }\href
  {\doibase 10.1103/PhysRevD.66.034506} {\bibfield  {journal} {\bibinfo
  {journal} {Phys. Rev.}\ }\textbf {\bibinfo {volume} {D66}},\ \bibinfo {pages}
  {034506} (\bibinfo {year} {2002})},\ \Eprint
  {http://arxiv.org/abs/hep-lat/0201021} {arXiv:hep-lat/0201021 [hep-lat]}
  \BibitemShut {NoStop}%
\bibitem [{\citenamefont {Ball}\ \emph {et~al.}(2015)\citenamefont {Ball} \emph
  {et~al.}}]{Ball:2014uwa}%
  \BibitemOpen
  \bibfield  {author} {\bibinfo {author} {\bibfnamefont {R.~D.}\ \bibnamefont
  {Ball}} \emph {et~al.} (\bibinfo {collaboration} {NNPDF}),\ }\href@noop {}
  {\bibfield  {journal} {\bibinfo  {journal} {JHEP}\ }\textbf {\bibinfo
  {volume} {04}},\ \bibinfo {pages} {040} (\bibinfo {year} {2015})},\ \Eprint
  {http://arxiv.org/abs/1410.8849} {arXiv:1410.8849 [hep-ph]} \BibitemShut
  {NoStop}%
\bibitem [{\citenamefont {Dulat}\ \emph {et~al.}(2016)\citenamefont {Dulat},
  \citenamefont {Hou}, \citenamefont {Gao}, \citenamefont {Guzzi},
  \citenamefont {Huston}, \citenamefont {Nadolsky}, \citenamefont {Pumplin},
  \citenamefont {Schmidt}, \citenamefont {Stump},\ and\ \citenamefont
  {Yuan}}]{Dulat:2015mca}%
  \BibitemOpen
  \bibfield  {author} {\bibinfo {author} {\bibfnamefont {S.}~\bibnamefont
  {Dulat}}, \bibinfo {author} {\bibfnamefont {T.-J.}\ \bibnamefont {Hou}},
  \bibinfo {author} {\bibfnamefont {J.}~\bibnamefont {Gao}}, \bibinfo {author}
  {\bibfnamefont {M.}~\bibnamefont {Guzzi}}, \bibinfo {author} {\bibfnamefont
  {J.}~\bibnamefont {Huston}}, \bibinfo {author} {\bibfnamefont
  {P.}~\bibnamefont {Nadolsky}}, \bibinfo {author} {\bibfnamefont
  {J.}~\bibnamefont {Pumplin}}, \bibinfo {author} {\bibfnamefont
  {C.}~\bibnamefont {Schmidt}}, \bibinfo {author} {\bibfnamefont
  {D.}~\bibnamefont {Stump}}, \ and\ \bibinfo {author} {\bibfnamefont {C.~P.}\
  \bibnamefont {Yuan}},\ }\href@noop {} {\bibfield  {journal} {\bibinfo
  {journal} {Phys. Rev.}\ }\textbf {\bibinfo {volume} {D93}},\ \bibinfo {pages}
  {033006} (\bibinfo {year} {2016})},\ \Eprint
  {http://arxiv.org/abs/1506.07443} {arXiv:1506.07443 [hep-ph]} \BibitemShut
  {NoStop}%
\bibitem [{\citenamefont {Martin}\ \emph {et~al.}(2009)\citenamefont {Martin},
  \citenamefont {Stirling}, \citenamefont {Thorne},\ and\ \citenamefont
  {Watt}}]{Martin:2009iq}%
  \BibitemOpen
  \bibfield  {author} {\bibinfo {author} {\bibfnamefont {A.~D.}\ \bibnamefont
  {Martin}}, \bibinfo {author} {\bibfnamefont {W.~J.}\ \bibnamefont
  {Stirling}}, \bibinfo {author} {\bibfnamefont {R.~S.}\ \bibnamefont
  {Thorne}}, \ and\ \bibinfo {author} {\bibfnamefont {G.}~\bibnamefont
  {Watt}},\ }\href@noop {} {\bibfield  {journal} {\bibinfo  {journal} {Eur.
  Phys. J.}\ }\textbf {\bibinfo {volume} {C63}},\ \bibinfo {pages} {189}
  (\bibinfo {year} {2009})},\ \Eprint {http://arxiv.org/abs/0901.0002}
  {arXiv:0901.0002 [hep-ph]} \BibitemShut {NoStop}%
\bibitem [{\citenamefont {Alekhin}\ \emph {et~al.}(2017)\citenamefont
  {Alekhin}, \citenamefont {Bl{\"u}mlein}, \citenamefont {Moch},\ and\
  \citenamefont {Placakyte}}]{Alekhin:2017kpj}%
  \BibitemOpen
  \bibfield  {author} {\bibinfo {author} {\bibfnamefont {S.}~\bibnamefont
  {Alekhin}}, \bibinfo {author} {\bibfnamefont {J.}~\bibnamefont
  {Bl{\"u}mlein}}, \bibinfo {author} {\bibfnamefont {S.}~\bibnamefont {Moch}},
  \ and\ \bibinfo {author} {\bibfnamefont {R.}~\bibnamefont {Placakyte}},\
  }\href@noop {} {\  (\bibinfo {year} {2017})},\ \Eprint
  {http://arxiv.org/abs/1701.05838} {arXiv:1701.05838 [hep-ph]} \BibitemShut
  {NoStop}%
\bibitem [{\citenamefont {Buckley}\ \emph {et~al.}(2015)\citenamefont
  {Buckley}, \citenamefont {Ferrando}, \citenamefont {Lloyd}, \citenamefont
  {Nordstr{\"o}m}, \citenamefont {Page}, \citenamefont {R{\"u}fenacht},
  \citenamefont {Sch{\"o}nherr},\ and\ \citenamefont {Watt}}]{Buckley:2014ana}%
  \BibitemOpen
  \bibfield  {author} {\bibinfo {author} {\bibfnamefont {A.}~\bibnamefont
  {Buckley}}, \bibinfo {author} {\bibfnamefont {J.}~\bibnamefont {Ferrando}},
  \bibinfo {author} {\bibfnamefont {S.}~\bibnamefont {Lloyd}}, \bibinfo
  {author} {\bibfnamefont {K.}~\bibnamefont {Nordstr{\"o}m}}, \bibinfo {author}
  {\bibfnamefont {B.}~\bibnamefont {Page}}, \bibinfo {author} {\bibfnamefont
  {M.}~\bibnamefont {R{\"u}fenacht}}, \bibinfo {author} {\bibfnamefont
  {M.}~\bibnamefont {Sch{\"o}nherr}}, \ and\ \bibinfo {author} {\bibfnamefont
  {G.}~\bibnamefont {Watt}},\ }\href@noop {} {\bibfield  {journal} {\bibinfo
  {journal} {Eur. Phys. J.}\ }\textbf {\bibinfo {volume} {C75}},\ \bibinfo
  {pages} {132} (\bibinfo {year} {2015})},\ \Eprint
  {http://arxiv.org/abs/1412.7420} {arXiv:1412.7420 [hep-ph]} \BibitemShut
  {NoStop}%
\bibitem [{\citenamefont {Ji}(2013)}]{Ji:2013dva}%
  \BibitemOpen
  \bibfield  {author} {\bibinfo {author} {\bibfnamefont {X.}~\bibnamefont
  {Ji}},\ }\href {\doibase 10.1103/PhysRevLett.110.262002} {\bibfield
  {journal} {\bibinfo  {journal} {Phys. Rev. Lett.}\ }\textbf {\bibinfo
  {volume} {110}},\ \bibinfo {pages} {262002} (\bibinfo {year} {2013})},\
  \Eprint {http://arxiv.org/abs/1305.1539} {arXiv:1305.1539 [hep-ph]}
  \BibitemShut {NoStop}%
\bibitem [{\citenamefont {Stewart}\ \emph
  {et~al.}(2010{\natexlab{a}})\citenamefont {Stewart}, \citenamefont
  {Tackmann},\ and\ \citenamefont {Waalewijn}}]{Stewart:2009yx}%
  \BibitemOpen
  \bibfield  {author} {\bibinfo {author} {\bibfnamefont {I.~W.}\ \bibnamefont
  {Stewart}}, \bibinfo {author} {\bibfnamefont {F.~J.}\ \bibnamefont
  {Tackmann}}, \ and\ \bibinfo {author} {\bibfnamefont {W.~J.}\ \bibnamefont
  {Waalewijn}},\ }\href {\doibase 10.1103/PhysRevD.81.094035} {\bibfield
  {journal} {\bibinfo  {journal} {Phys.Rev.}\ }\textbf {\bibinfo {volume}
  {D81}},\ \bibinfo {pages} {094035} (\bibinfo {year} {2010}{\natexlab{a}})},\
  \Eprint {http://arxiv.org/abs/0910.0467} {arXiv:0910.0467 [hep-ph]}
  \BibitemShut {NoStop}%
\bibitem [{\citenamefont {Stewart}\ \emph
  {et~al.}(2010{\natexlab{b}})\citenamefont {Stewart}, \citenamefont
  {Tackmann},\ and\ \citenamefont {Waalewijn}}]{Stewart:2010qs}%
  \BibitemOpen
  \bibfield  {author} {\bibinfo {author} {\bibfnamefont {I.~W.}\ \bibnamefont
  {Stewart}}, \bibinfo {author} {\bibfnamefont {F.~J.}\ \bibnamefont
  {Tackmann}}, \ and\ \bibinfo {author} {\bibfnamefont {W.~J.}\ \bibnamefont
  {Waalewijn}},\ }\href {\doibase 10.1007/JHEP09(2010)005} {\bibfield
  {journal} {\bibinfo  {journal} {JHEP}\ }\textbf {\bibinfo {volume} {1009}},\
  \bibinfo {pages} {005} (\bibinfo {year} {2010}{\natexlab{b}})},\ \Eprint
  {http://arxiv.org/abs/1002.2213} {arXiv:1002.2213 [hep-ph]} \BibitemShut
  {NoStop}%
\bibitem [{\citenamefont {Ji}\ and\ \citenamefont {Zhang}(2015)}]{Ji:2015jwa}%
  \BibitemOpen
  \bibfield  {author} {\bibinfo {author} {\bibfnamefont {X.}~\bibnamefont
  {Ji}}\ and\ \bibinfo {author} {\bibfnamefont {J.-H.}\ \bibnamefont {Zhang}},\
  }\href@noop {} {\bibfield  {journal} {\bibinfo  {journal} {Phys. Rev.}\
  }\textbf {\bibinfo {volume} {D92}},\ \bibinfo {pages} {034006} (\bibinfo
  {year} {2015})},\ \Eprint {http://arxiv.org/abs/1505.07699} {arXiv:1505.07699
  [hep-ph]} \BibitemShut {NoStop}%
\bibitem [{\citenamefont {Ji}\ \emph {et~al.}(2018)\citenamefont {Ji},
  \citenamefont {Zhang},\ and\ \citenamefont {Zhao}}]{Ji:2017oey}%
  \BibitemOpen
  \bibfield  {author} {\bibinfo {author} {\bibfnamefont {X.}~\bibnamefont
  {Ji}}, \bibinfo {author} {\bibfnamefont {J.-H.}\ \bibnamefont {Zhang}}, \
  and\ \bibinfo {author} {\bibfnamefont {Y.}~\bibnamefont {Zhao}},\ }\href
  {\doibase 10.1103/PhysRevLett.120.112001} {\bibfield  {journal} {\bibinfo
  {journal} {Phys. Rev. Lett.}\ }\textbf {\bibinfo {volume} {120}},\ \bibinfo
  {pages} {112001} (\bibinfo {year} {2018})},\ \Eprint
  {http://arxiv.org/abs/1706.08962} {arXiv:1706.08962 [hep-ph]} \BibitemShut
  {NoStop}%
\bibitem [{\citenamefont {Ishikawa}\ \emph {et~al.}(2017)\citenamefont
  {Ishikawa}, \citenamefont {Ma}, \citenamefont {Qiu},\ and\ \citenamefont
  {Yoshida}}]{Ishikawa:2017faj}%
  \BibitemOpen
  \bibfield  {author} {\bibinfo {author} {\bibfnamefont {T.}~\bibnamefont
  {Ishikawa}}, \bibinfo {author} {\bibfnamefont {Y.-Q.}\ \bibnamefont {Ma}},
  \bibinfo {author} {\bibfnamefont {J.-W.}\ \bibnamefont {Qiu}}, \ and\
  \bibinfo {author} {\bibfnamefont {S.}~\bibnamefont {Yoshida}},\ }\href@noop
  {} {\  (\bibinfo {year} {2017})},\ \Eprint {http://arxiv.org/abs/1707.03107}
  {arXiv:1707.03107 [hep-ph]} \BibitemShut {NoStop}%
\bibitem [{\citenamefont {Ji}(2014)}]{Ji:2014gla}%
  \BibitemOpen
  \bibfield  {author} {\bibinfo {author} {\bibfnamefont {X.}~\bibnamefont
  {Ji}},\ }\href@noop {} {\bibfield  {journal} {\bibinfo  {journal} {Sci. China
  Phys. Mech. Astron.}\ }\textbf {\bibinfo {volume} {57}},\ \bibinfo {pages}
  {1407} (\bibinfo {year} {2014})},\ \Eprint {http://arxiv.org/abs/1404.6680}
  {arXiv:1404.6680 [hep-ph]} \BibitemShut {NoStop}%
\bibitem [{\citenamefont {Izubuchi}\ \emph {et~al.}(2018)\citenamefont
  {Izubuchi}, \citenamefont {Ji}, \citenamefont {Jin}, \citenamefont
  {Stewart},\ and\ \citenamefont {Zhao}}]{Izubuchi:2018srq}%
  \BibitemOpen
  \bibfield  {author} {\bibinfo {author} {\bibfnamefont {T.}~\bibnamefont
  {Izubuchi}}, \bibinfo {author} {\bibfnamefont {X.}~\bibnamefont {Ji}},
  \bibinfo {author} {\bibfnamefont {L.}~\bibnamefont {Jin}}, \bibinfo {author}
  {\bibfnamefont {I.~W.}\ \bibnamefont {Stewart}}, \ and\ \bibinfo {author}
  {\bibfnamefont {Y.}~\bibnamefont {Zhao}},\ }\href@noop {} {\  (\bibinfo
  {year} {2018})},\ \Eprint {http://arxiv.org/abs/1801.03917} {arXiv:1801.03917
  [hep-ph]} \BibitemShut {NoStop}%
\bibitem [{\citenamefont {Lin}\ \emph {et~al.}(2015)\citenamefont {Lin},
  \citenamefont {Chen}, \citenamefont {Cohen},\ and\ \citenamefont
  {Ji}}]{Lin:2014zya}%
  \BibitemOpen
  \bibfield  {author} {\bibinfo {author} {\bibfnamefont {H.-W.}\ \bibnamefont
  {Lin}}, \bibinfo {author} {\bibfnamefont {J.-W.}\ \bibnamefont {Chen}},
  \bibinfo {author} {\bibfnamefont {S.~D.}\ \bibnamefont {Cohen}}, \ and\
  \bibinfo {author} {\bibfnamefont {X.}~\bibnamefont {Ji}},\ }\href@noop {}
  {\bibfield  {journal} {\bibinfo  {journal} {Phys. Rev.}\ }\textbf {\bibinfo
  {volume} {D91}},\ \bibinfo {pages} {054510} (\bibinfo {year} {2015})},\
  \Eprint {http://arxiv.org/abs/1402.1462} {arXiv:1402.1462 [hep-ph]}
  \BibitemShut {NoStop}%
\bibitem [{\citenamefont {Alexandrou}\ \emph {et~al.}(2015)\citenamefont
  {Alexandrou}, \citenamefont {Cichy}, \citenamefont {Drach}, \citenamefont
  {Garcia-Ramos}, \citenamefont {Hadjiyiannakou}, \citenamefont {Jansen},
  \citenamefont {Steffens},\ and\ \citenamefont {Wiese}}]{Alexandrou:2015rja}%
  \BibitemOpen
  \bibfield  {author} {\bibinfo {author} {\bibfnamefont {C.}~\bibnamefont
  {Alexandrou}}, \bibinfo {author} {\bibfnamefont {K.}~\bibnamefont {Cichy}},
  \bibinfo {author} {\bibfnamefont {V.}~\bibnamefont {Drach}}, \bibinfo
  {author} {\bibfnamefont {E.}~\bibnamefont {Garcia-Ramos}}, \bibinfo {author}
  {\bibfnamefont {K.}~\bibnamefont {Hadjiyiannakou}}, \bibinfo {author}
  {\bibfnamefont {K.}~\bibnamefont {Jansen}}, \bibinfo {author} {\bibfnamefont
  {F.}~\bibnamefont {Steffens}}, \ and\ \bibinfo {author} {\bibfnamefont
  {C.}~\bibnamefont {Wiese}},\ }\href@noop {} {\bibfield  {journal} {\bibinfo
  {journal} {Phys. Rev.}\ }\textbf {\bibinfo {volume} {D92}},\ \bibinfo {pages}
  {014502} (\bibinfo {year} {2015})},\ \Eprint
  {http://arxiv.org/abs/1504.07455} {arXiv:1504.07455 [hep-lat]} \BibitemShut
  {NoStop}%
\bibitem [{\citenamefont {Chen}\ \emph
  {et~al.}(2016{\natexlab{a}})\citenamefont {Chen}, \citenamefont {Cohen},
  \citenamefont {Ji}, \citenamefont {Lin},\ and\ \citenamefont
  {Zhang}}]{Chen:2016utp}%
  \BibitemOpen
  \bibfield  {author} {\bibinfo {author} {\bibfnamefont {J.-W.}\ \bibnamefont
  {Chen}}, \bibinfo {author} {\bibfnamefont {S.~D.}\ \bibnamefont {Cohen}},
  \bibinfo {author} {\bibfnamefont {X.}~\bibnamefont {Ji}}, \bibinfo {author}
  {\bibfnamefont {H.-W.}\ \bibnamefont {Lin}}, \ and\ \bibinfo {author}
  {\bibfnamefont {J.-H.}\ \bibnamefont {Zhang}},\ }\href@noop {} {\bibfield
  {journal} {\bibinfo  {journal} {Nucl. Phys.}\ }\textbf {\bibinfo {volume}
  {B911}},\ \bibinfo {pages} {246} (\bibinfo {year} {2016}{\natexlab{a}})},\
  \Eprint {http://arxiv.org/abs/1603.06664} {arXiv:1603.06664 [hep-ph]}
  \BibitemShut {NoStop}%
\bibitem [{\citenamefont {Alexandrou}\ \emph {et~al.}(2016)\citenamefont
  {Alexandrou}, \citenamefont {Cichy}, \citenamefont {Constantinou},
  \citenamefont {Hadjiyiannakou}, \citenamefont {Jansen}, \citenamefont
  {Steffens},\ and\ \citenamefont {Wiese}}]{Alexandrou:2016jqi}%
  \BibitemOpen
  \bibfield  {author} {\bibinfo {author} {\bibfnamefont {C.}~\bibnamefont
  {Alexandrou}}, \bibinfo {author} {\bibfnamefont {K.}~\bibnamefont {Cichy}},
  \bibinfo {author} {\bibfnamefont {M.}~\bibnamefont {Constantinou}}, \bibinfo
  {author} {\bibfnamefont {K.}~\bibnamefont {Hadjiyiannakou}}, \bibinfo
  {author} {\bibfnamefont {K.}~\bibnamefont {Jansen}}, \bibinfo {author}
  {\bibfnamefont {F.}~\bibnamefont {Steffens}}, \ and\ \bibinfo {author}
  {\bibfnamefont {C.}~\bibnamefont {Wiese}},\ }\href@noop {} {\  (\bibinfo
  {year} {2016})},\ \Eprint {http://arxiv.org/abs/1610.03689} {arXiv:1610.03689
  [hep-lat]} \BibitemShut {NoStop}%
\bibitem [{\citenamefont {Zhang}\ \emph {et~al.}(2017)\citenamefont {Zhang},
  \citenamefont {Chen}, \citenamefont {Ji}, \citenamefont {Jin},\ and\
  \citenamefont {Lin}}]{Zhang:2017bzy}%
  \BibitemOpen
  \bibfield  {author} {\bibinfo {author} {\bibfnamefont {J.-H.}\ \bibnamefont
  {Zhang}}, \bibinfo {author} {\bibfnamefont {J.-W.}\ \bibnamefont {Chen}},
  \bibinfo {author} {\bibfnamefont {X.}~\bibnamefont {Ji}}, \bibinfo {author}
  {\bibfnamefont {L.}~\bibnamefont {Jin}}, \ and\ \bibinfo {author}
  {\bibfnamefont {H.-W.}\ \bibnamefont {Lin}},\ }\href@noop {} {\bibfield
  {journal} {\bibinfo  {journal} {Phys. Rev.}\ }\textbf {\bibinfo {volume}
  {D95}},\ \bibinfo {pages} {094514} (\bibinfo {year} {2017})},\ \Eprint
  {http://arxiv.org/abs/1702.00008} {arXiv:1702.00008 [hep-lat]} \BibitemShut
  {NoStop}%
\bibitem [{\citenamefont {Alexandrou}\ \emph {et~al.}(2017)\citenamefont
  {Alexandrou}, \citenamefont {Cichy}, \citenamefont {Constantinou},
  \citenamefont {Hadjiyiannakou}, \citenamefont {Jansen}, \citenamefont
  {Panagopoulos},\ and\ \citenamefont {Steffens}}]{Alexandrou:2017huk}%
  \BibitemOpen
  \bibfield  {author} {\bibinfo {author} {\bibfnamefont {C.}~\bibnamefont
  {Alexandrou}}, \bibinfo {author} {\bibfnamefont {K.}~\bibnamefont {Cichy}},
  \bibinfo {author} {\bibfnamefont {M.}~\bibnamefont {Constantinou}}, \bibinfo
  {author} {\bibfnamefont {K.}~\bibnamefont {Hadjiyiannakou}}, \bibinfo
  {author} {\bibfnamefont {K.}~\bibnamefont {Jansen}}, \bibinfo {author}
  {\bibfnamefont {H.}~\bibnamefont {Panagopoulos}}, \ and\ \bibinfo {author}
  {\bibfnamefont {F.}~\bibnamefont {Steffens}},\ }\href@noop {} {\  (\bibinfo
  {year} {2017})},\ \Eprint {http://arxiv.org/abs/1706.00265} {arXiv:1706.00265
  [hep-lat]} \BibitemShut {NoStop}%
\bibitem [{\citenamefont {Chen}\ \emph {et~al.}(2017)\citenamefont {Chen},
  \citenamefont {Ishikawa}, \citenamefont {Jin}, \citenamefont {Lin},
  \citenamefont {Yang}, \citenamefont {Zhang},\ and\ \citenamefont
  {Zhao}}]{Chen:2017mzz}%
  \BibitemOpen
  \bibfield  {author} {\bibinfo {author} {\bibfnamefont {J.-W.}\ \bibnamefont
  {Chen}}, \bibinfo {author} {\bibfnamefont {T.}~\bibnamefont {Ishikawa}},
  \bibinfo {author} {\bibfnamefont {L.}~\bibnamefont {Jin}}, \bibinfo {author}
  {\bibfnamefont {H.-W.}\ \bibnamefont {Lin}}, \bibinfo {author} {\bibfnamefont
  {Y.-B.}\ \bibnamefont {Yang}}, \bibinfo {author} {\bibfnamefont {J.-H.}\
  \bibnamefont {Zhang}}, \ and\ \bibinfo {author} {\bibfnamefont
  {Y.}~\bibnamefont {Zhao}},\ }\href@noop {} {\  (\bibinfo {year} {2017})},\
  \Eprint {http://arxiv.org/abs/1706.01295} {arXiv:1706.01295 [hep-lat]}
  \BibitemShut {NoStop}%
\bibitem [{\citenamefont {Lin}\ \emph {et~al.}(2017)\citenamefont {Lin},
  \citenamefont {Chen}, \citenamefont {Ishikawa},\ and\ \citenamefont
  {Zhang}}]{Lin:2017ani}%
  \BibitemOpen
  \bibfield  {author} {\bibinfo {author} {\bibfnamefont {H.-W.}\ \bibnamefont
  {Lin}}, \bibinfo {author} {\bibfnamefont {J.-W.}\ \bibnamefont {Chen}},
  \bibinfo {author} {\bibfnamefont {T.}~\bibnamefont {Ishikawa}}, \ and\
  \bibinfo {author} {\bibfnamefont {J.-H.}\ \bibnamefont {Zhang}},\ }\href@noop
  {} {\  (\bibinfo {year} {2017})},\ \Eprint {http://arxiv.org/abs/1708.05301}
  {arXiv:1708.05301 [hep-lat]} \BibitemShut {NoStop}%
\bibitem [{\citenamefont {Xiong}\ \emph {et~al.}(2014)\citenamefont {Xiong},
  \citenamefont {Ji}, \citenamefont {Zhang},\ and\ \citenamefont
  {Zhao}}]{Xiong:2013bka}%
  \BibitemOpen
  \bibfield  {author} {\bibinfo {author} {\bibfnamefont {X.}~\bibnamefont
  {Xiong}}, \bibinfo {author} {\bibfnamefont {X.}~\bibnamefont {Ji}}, \bibinfo
  {author} {\bibfnamefont {J.-H.}\ \bibnamefont {Zhang}}, \ and\ \bibinfo
  {author} {\bibfnamefont {Y.}~\bibnamefont {Zhao}},\ }\href@noop {} {\bibfield
   {journal} {\bibinfo  {journal} {Phys. Rev.}\ }\textbf {\bibinfo {volume}
  {D90}},\ \bibinfo {pages} {014051} (\bibinfo {year} {2014})},\ \Eprint
  {http://arxiv.org/abs/1310.7471} {arXiv:1310.7471 [hep-ph]} \BibitemShut
  {NoStop}%
\bibitem [{\citenamefont {Ma}\ and\ \citenamefont {Qiu}(2014)}]{Ma:2014jla}%
  \BibitemOpen
  \bibfield  {author} {\bibinfo {author} {\bibfnamefont {Y.-Q.}\ \bibnamefont
  {Ma}}\ and\ \bibinfo {author} {\bibfnamefont {J.-W.}\ \bibnamefont {Qiu}},\
  }\href@noop {} {\  (\bibinfo {year} {2014})},\ \Eprint
  {http://arxiv.org/abs/1404.6860} {arXiv:1404.6860 [hep-ph]} \BibitemShut
  {NoStop}%
\bibitem [{\citenamefont {Ishikawa}\ \emph {et~al.}(2016)\citenamefont
  {Ishikawa}, \citenamefont {Ma}, \citenamefont {Qiu},\ and\ \citenamefont
  {Yoshida}}]{Ishikawa:2016znu}%
  \BibitemOpen
  \bibfield  {author} {\bibinfo {author} {\bibfnamefont {T.}~\bibnamefont
  {Ishikawa}}, \bibinfo {author} {\bibfnamefont {Y.-Q.}\ \bibnamefont {Ma}},
  \bibinfo {author} {\bibfnamefont {J.-W.}\ \bibnamefont {Qiu}}, \ and\
  \bibinfo {author} {\bibfnamefont {S.}~\bibnamefont {Yoshida}},\ }\href@noop
  {} {\  (\bibinfo {year} {2016})},\ \Eprint {http://arxiv.org/abs/1609.02018}
  {arXiv:1609.02018 [hep-lat]} \BibitemShut {NoStop}%
\bibitem [{\citenamefont {Chen}\ \emph
  {et~al.}(2016{\natexlab{b}})\citenamefont {Chen}, \citenamefont {Ji},\ and\
  \citenamefont {Zhang}}]{Chen:2016fxx}%
  \BibitemOpen
  \bibfield  {author} {\bibinfo {author} {\bibfnamefont {J.-W.}\ \bibnamefont
  {Chen}}, \bibinfo {author} {\bibfnamefont {X.}~\bibnamefont {Ji}}, \ and\
  \bibinfo {author} {\bibfnamefont {J.-H.}\ \bibnamefont {Zhang}},\ }\href@noop
  {} {\  (\bibinfo {year} {2016}{\natexlab{b}})},\ \Eprint
  {http://arxiv.org/abs/1609.08102} {arXiv:1609.08102 [hep-ph]} \BibitemShut
  {NoStop}%
\bibitem [{\citenamefont {Xiong}\ \emph {et~al.}(2017)\citenamefont {Xiong},
  \citenamefont {Luu},\ and\ \citenamefont {Mei{\ss}ner}}]{Xiong:2017jtn}%
  \BibitemOpen
  \bibfield  {author} {\bibinfo {author} {\bibfnamefont {X.}~\bibnamefont
  {Xiong}}, \bibinfo {author} {\bibfnamefont {T.}~\bibnamefont {Luu}}, \ and\
  \bibinfo {author} {\bibfnamefont {U.-G.}\ \bibnamefont {Mei{\ss}ner}},\
  }\href@noop {} {\  (\bibinfo {year} {2017})},\ \Eprint
  {http://arxiv.org/abs/1705.00246} {arXiv:1705.00246 [hep-ph]} \BibitemShut
  {NoStop}%
\bibitem [{\citenamefont {Constantinou}\ and\ \citenamefont
  {Panagopoulos}(2017)}]{Constantinou:2017sej}%
  \BibitemOpen
  \bibfield  {author} {\bibinfo {author} {\bibfnamefont {M.}~\bibnamefont
  {Constantinou}}\ and\ \bibinfo {author} {\bibfnamefont {H.}~\bibnamefont
  {Panagopoulos}},\ }\href@noop {} {\  (\bibinfo {year} {2017})},\ \Eprint
  {http://arxiv.org/abs/1705.11193} {arXiv:1705.11193 [hep-lat]} \BibitemShut
  {NoStop}%
\bibitem [{\citenamefont {Green}\ \emph {et~al.}(2017)\citenamefont {Green},
  \citenamefont {Jansen},\ and\ \citenamefont {Steffens}}]{Green:2017xeu}%
  \BibitemOpen
  \bibfield  {author} {\bibinfo {author} {\bibfnamefont {J.}~\bibnamefont
  {Green}}, \bibinfo {author} {\bibfnamefont {K.}~\bibnamefont {Jansen}}, \
  and\ \bibinfo {author} {\bibfnamefont {F.}~\bibnamefont {Steffens}},\
  }\href@noop {} {\  (\bibinfo {year} {2017})},\ \Eprint
  {http://arxiv.org/abs/1707.07152} {arXiv:1707.07152 [hep-lat]} \BibitemShut
  {NoStop}%
\bibitem [{\citenamefont {Radyushkin}(2017)}]{Radyushkin:2017cyf}%
  \BibitemOpen
  \bibfield  {author} {\bibinfo {author} {\bibfnamefont {A.~V.}\ \bibnamefont
  {Radyushkin}},\ }\href {\doibase 10.1103/PhysRevD.96.034025} {\bibfield
  {journal} {\bibinfo  {journal} {Phys. Rev.}\ }\textbf {\bibinfo {volume}
  {D96}},\ \bibinfo {pages} {034025} (\bibinfo {year} {2017})},\ \Eprint
  {http://arxiv.org/abs/1705.01488} {arXiv:1705.01488 [hep-ph]} \BibitemShut
  {NoStop}%
\bibitem [{\citenamefont {Orginos}\ \emph {et~al.}(2017)\citenamefont
  {Orginos}, \citenamefont {Radyushkin}, \citenamefont {Karpie},\ and\
  \citenamefont {Zafeiropoulos}}]{Orginos:2017kos}%
  \BibitemOpen
  \bibfield  {author} {\bibinfo {author} {\bibfnamefont {K.}~\bibnamefont
  {Orginos}}, \bibinfo {author} {\bibfnamefont {A.}~\bibnamefont {Radyushkin}},
  \bibinfo {author} {\bibfnamefont {J.}~\bibnamefont {Karpie}}, \ and\ \bibinfo
  {author} {\bibfnamefont {S.}~\bibnamefont {Zafeiropoulos}},\ }\href@noop {}
  {\  (\bibinfo {year} {2017})},\ \Eprint {http://arxiv.org/abs/1706.05373}
  {arXiv:1706.05373 [hep-ph]} \BibitemShut {NoStop}%
\bibitem [{\citenamefont {Capitani}(2003)}]{Capitani:2002mp}%
  \BibitemOpen
  \bibfield  {author} {\bibinfo {author} {\bibfnamefont {S.}~\bibnamefont
  {Capitani}},\ }\href@noop {} {\bibfield  {journal} {\bibinfo  {journal}
  {Phys. Rept.}\ }\textbf {\bibinfo {volume} {382}},\ \bibinfo {pages} {113}
  (\bibinfo {year} {2003})},\ \Eprint {http://arxiv.org/abs/hep-lat/0211036}
  {arXiv:hep-lat/0211036 [hep-lat]} \BibitemShut {NoStop}%
\bibitem [{\citenamefont {Martinelli}\ \emph {et~al.}(1995)\citenamefont
  {Martinelli}, \citenamefont {Pittori}, \citenamefont {Sachrajda},
  \citenamefont {Testa},\ and\ \citenamefont {Vladikas}}]{Martinelli:1994ty}%
  \BibitemOpen
  \bibfield  {author} {\bibinfo {author} {\bibfnamefont {G.}~\bibnamefont
  {Martinelli}}, \bibinfo {author} {\bibfnamefont {C.}~\bibnamefont {Pittori}},
  \bibinfo {author} {\bibfnamefont {C.~T.}\ \bibnamefont {Sachrajda}}, \bibinfo
  {author} {\bibfnamefont {M.}~\bibnamefont {Testa}}, \ and\ \bibinfo {author}
  {\bibfnamefont {A.}~\bibnamefont {Vladikas}},\ }\href@noop {} {\bibfield
  {journal} {\bibinfo  {journal} {Nucl. Phys.}\ }\textbf {\bibinfo {volume}
  {B445}},\ \bibinfo {pages} {81} (\bibinfo {year} {1995})},\ \Eprint
  {http://arxiv.org/abs/hep-lat/9411010} {arXiv:hep-lat/9411010 [hep-lat]}
  \BibitemShut {NoStop}%
\bibitem [{\citenamefont {Carlson}\ and\ \citenamefont
  {Freid}(2017)}]{Carlson:2017gpk}%
  \BibitemOpen
  \bibfield  {author} {\bibinfo {author} {\bibfnamefont {C.~E.}\ \bibnamefont
  {Carlson}}\ and\ \bibinfo {author} {\bibfnamefont {M.}~\bibnamefont
  {Freid}},\ }\href@noop {} {\  (\bibinfo {year} {2017})},\ \Eprint
  {http://arxiv.org/abs/1702.05775} {arXiv:1702.05775 [hep-ph]} \BibitemShut
  {NoStop}%
\bibitem [{\citenamefont {Brice{\~n}o}\ \emph {et~al.}(2017)\citenamefont
  {Brice{\~n}o}, \citenamefont {Hansen},\ and\ \citenamefont
  {Monahan}}]{Briceno:2017cpo}%
  \BibitemOpen
  \bibfield  {author} {\bibinfo {author} {\bibfnamefont {R.~A.}\ \bibnamefont
  {Brice{\~n}o}}, \bibinfo {author} {\bibfnamefont {M.~T.}\ \bibnamefont
  {Hansen}}, \ and\ \bibinfo {author} {\bibfnamefont {C.~J.}\ \bibnamefont
  {Monahan}},\ }\href@noop {} {\  (\bibinfo {year} {2017})},\ \Eprint
  {http://arxiv.org/abs/1703.06072} {arXiv:1703.06072 [hep-lat]} \BibitemShut
  {NoStop}%
\bibitem [{\citenamefont {Ji}\ \emph {et~al.}(2017)\citenamefont {Ji},
  \citenamefont {Zhang},\ and\ \citenamefont {Zhao}}]{Ji:2017rah}%
  \BibitemOpen
  \bibfield  {author} {\bibinfo {author} {\bibfnamefont {X.}~\bibnamefont
  {Ji}}, \bibinfo {author} {\bibfnamefont {J.-H.}\ \bibnamefont {Zhang}}, \
  and\ \bibinfo {author} {\bibfnamefont {Y.}~\bibnamefont {Zhao}},\ }\href@noop
  {} {\  (\bibinfo {year} {2017})},\ \Eprint {http://arxiv.org/abs/1706.07416}
  {arXiv:1706.07416 [hep-ph]} \BibitemShut {NoStop}%
\bibitem [{\citenamefont {Dotsenko}\ and\ \citenamefont
  {Vergeles}(1980)}]{Dotsenko:1979wb}%
  \BibitemOpen
  \bibfield  {author} {\bibinfo {author} {\bibfnamefont {V.~S.}\ \bibnamefont
  {Dotsenko}}\ and\ \bibinfo {author} {\bibfnamefont {S.~N.}\ \bibnamefont
  {Vergeles}},\ }\href@noop {} {\bibfield  {journal} {\bibinfo  {journal}
  {Nucl. Phys.}\ }\textbf {\bibinfo {volume} {B169}},\ \bibinfo {pages} {527}
  (\bibinfo {year} {1980})}\BibitemShut {NoStop}%
\bibitem [{\citenamefont {Craigie}\ and\ \citenamefont
  {Dorn}(1981)}]{Craigie:1980qs}%
  \BibitemOpen
  \bibfield  {author} {\bibinfo {author} {\bibfnamefont {N.~S.}\ \bibnamefont
  {Craigie}}\ and\ \bibinfo {author} {\bibfnamefont {H.}~\bibnamefont {Dorn}},\
  }\href@noop {} {\bibfield  {journal} {\bibinfo  {journal} {Nucl. Phys.}\
  }\textbf {\bibinfo {volume} {B185}},\ \bibinfo {pages} {204} (\bibinfo {year}
  {1981})}\BibitemShut {NoStop}%
\bibitem [{\citenamefont {Dorn}(1986)}]{Dorn:1986dt}%
  \BibitemOpen
  \bibfield  {author} {\bibinfo {author} {\bibfnamefont {H.}~\bibnamefont
  {Dorn}},\ }\href@noop {} {\bibfield  {journal} {\bibinfo  {journal} {Fortsch.
  Phys.}\ }\textbf {\bibinfo {volume} {34}},\ \bibinfo {pages} {11} (\bibinfo
  {year} {1986})}\BibitemShut {NoStop}%
\bibitem [{\citenamefont {Collins}(2013)}]{Collins:2011zzd}%
  \BibitemOpen
  \bibfield  {author} {\bibinfo {author} {\bibfnamefont {J.}~\bibnamefont
  {Collins}},\ }\href@noop {} {\emph {\bibinfo {title} {{Foundations of
  perturbative QCD}}}}\ (\bibinfo  {publisher} {Cambridge University Press},\
  \bibinfo {year} {2013})\BibitemShut {NoStop}%
\bibitem [{\citenamefont {Ji}\ \emph {et~al.}(2015)\citenamefont {Ji},
  \citenamefont {Sun}, \citenamefont {Xiong},\ and\ \citenamefont
  {Yuan}}]{Ji:2014hxa}%
  \BibitemOpen
  \bibfield  {author} {\bibinfo {author} {\bibfnamefont {X.}~\bibnamefont
  {Ji}}, \bibinfo {author} {\bibfnamefont {P.}~\bibnamefont {Sun}}, \bibinfo
  {author} {\bibfnamefont {X.}~\bibnamefont {Xiong}}, \ and\ \bibinfo {author}
  {\bibfnamefont {F.}~\bibnamefont {Yuan}},\ }\href@noop {} {\bibfield
  {journal} {\bibinfo  {journal} {Phys. Rev.}\ }\textbf {\bibinfo {volume}
  {D91}},\ \bibinfo {pages} {074009} (\bibinfo {year} {2015})},\ \Eprint
  {http://arxiv.org/abs/1405.7640} {arXiv:1405.7640 [hep-ph]} \BibitemShut
  {NoStop}%
\bibitem [{\citenamefont {Monahan}\ and\ \citenamefont
  {Orginos}(2016)}]{Monahan:2016bvm}%
  \BibitemOpen
  \bibfield  {author} {\bibinfo {author} {\bibfnamefont {C.}~\bibnamefont
  {Monahan}}\ and\ \bibinfo {author} {\bibfnamefont {K.}~\bibnamefont
  {Orginos}},\ }\href@noop {} {\  (\bibinfo {year} {2016})},\ \Eprint
  {http://arxiv.org/abs/1612.01584} {arXiv:1612.01584 [hep-lat]} \BibitemShut
  {NoStop}%
\bibitem [{\citenamefont {Li}(2016)}]{Li:2016amo}%
  \BibitemOpen
  \bibfield  {author} {\bibinfo {author} {\bibfnamefont {H.-n.}\ \bibnamefont
  {Li}},\ }\href {\doibase 10.1103/PhysRevD.94.074036} {\bibfield  {journal}
  {\bibinfo  {journal} {Phys. Rev.}\ }\textbf {\bibinfo {volume} {D94}},\
  \bibinfo {pages} {074036} (\bibinfo {year} {2016})},\ \Eprint
  {http://arxiv.org/abs/1602.07575} {arXiv:1602.07575 [hep-ph]} \BibitemShut
  {NoStop}%
\bibitem [{\citenamefont {Musch}\ \emph {et~al.}(2011)\citenamefont {Musch},
  \citenamefont {Hagler}, \citenamefont {Negele},\ and\ \citenamefont
  {Schafer}}]{Musch:2010ka}%
  \BibitemOpen
  \bibfield  {author} {\bibinfo {author} {\bibfnamefont {B.~U.}\ \bibnamefont
  {Musch}}, \bibinfo {author} {\bibfnamefont {P.}~\bibnamefont {Hagler}},
  \bibinfo {author} {\bibfnamefont {J.~W.}\ \bibnamefont {Negele}}, \ and\
  \bibinfo {author} {\bibfnamefont {A.}~\bibnamefont {Schafer}},\ }\href@noop
  {} {\bibfield  {journal} {\bibinfo  {journal} {Phys. Rev.}\ }\textbf
  {\bibinfo {volume} {D83}},\ \bibinfo {pages} {094507} (\bibinfo {year}
  {2011})},\ \Eprint {http://arxiv.org/abs/1011.1213} {arXiv:1011.1213
  [hep-lat]} \BibitemShut {NoStop}%
\bibitem [{\citenamefont {Rossi}\ and\ \citenamefont
  {Testa}(2017)}]{Rossi:2017muf}%
  \BibitemOpen
  \bibfield  {author} {\bibinfo {author} {\bibfnamefont {G.~C.}\ \bibnamefont
  {Rossi}}\ and\ \bibinfo {author} {\bibfnamefont {M.}~\bibnamefont {Testa}},\
  }\href {\doibase 10.1103/PhysRevD.96.014507} {\bibfield  {journal} {\bibinfo
  {journal} {Phys. Rev.}\ }\textbf {\bibinfo {volume} {D96}},\ \bibinfo {pages}
  {014507} (\bibinfo {year} {2017})},\ \Eprint
  {http://arxiv.org/abs/1706.04428} {arXiv:1706.04428 [hep-lat]} \BibitemShut
  {NoStop}%
\end{thebibliography}%

\end{document}